\documentclass[twocolumn,showpacs,preprintnumbers,amsmath,amssymb]{revtex4}

\usepackage{graphicx}
\usepackage{dcolumn}
\usepackage{bm}

\begin{document}

\title{Problem On The Acoustic Cloak By $0$ to $R_1$ Transformation
}

\author{Jianhua Li}
 \altaffiliation[Also at ]{GL Geophysical Laboratory, USA, glhua@glgeo.com}
\author{  Feng Xie, Lee Xie, Ganquan Xie}%
\affiliation{%
GL Geophysical Laboratory, USA
}%

\hfill\break
\author{Jianhua Li and Ganquan Xie}
\affiliation{
Chinese Dayuling Supercomputational Sciences Center, China \\
Hunan Super Computational Sciences Society, China
}%

\date{June 8, 2017}

\begin{abstract}
In this paper, we prove that for incident acoustic wave $p_i (\vec r)$ in outside of whole sphere $r_s \ge R_2 $, in the annular layer $R_1  \le R \le R_2 $, with anisotropic acoustic media induced by  0 to R1  spherical radial transformation,  pressure acoustic wave $P(\vec R)$, satisfies the anisotropic acoustic equation and no scattering interface condition on outer interface boundary $\vec R = \vec R_2 $, then on the inner interface boundary $\vec R = \vec R_1 $,$P(\vec R_1 )$ is nonzero constant, $P(\vec R_1 ^ +  ) \ne 0$, and the inner sphere $R\le R_1 $ can not be  cloaked, otherwise,  ¡°$P(\vec R) = 0$, in  $R \le R_1 $¡± that does cause contradiction between $P(\vec R_1 ^ +  ) \ne 0$ and $P(\vec R_1 ^ -  ) = 0$, $P(\vec R_1 ^ +  ) \ne P(\vec R_1 ^ -  )$, the contradiction shows that the nonzero  acoustic wave is propagation to penetrate the inner sphere, the inner sphere $R \le R_1$, can not be cloaked. Inversely, suppose that the inner sphere $R \le R_1$ is cloaked, the $P(\vec R)=0$ is the zero solution of acoustic equation in inner sphere, if the pressure acoustic wave $P(\vec R)$, satisfies the anisotropic  acoustic equation in annular layer $R_1  \le R \le R_2 $, and satisfies zero interface continuous conditions on the inner interface boundary $\vec R = \vec R_1 $, then on the outer interface boundary $\vec R = \vec R_2 $, $P(\vec R_2 ) \ne p_i (R_2 )$ ,the pressure wave is not equal to incident wave, which is contradiction with no scattering interface continuous condition on outer interface boundary $\vec R = \vec R_2 $, then there exist scattering wave to disturb the incident wave in outside of the whole sphere, $R \le R_2$, the whole sphere will be detected and exposed. The above two basic contraditions are caused by the inconsistent between induced anisotropic acoustic media in annular layer $R_1 \le R \le R_2$  by  0 to $R_1$ spherical radial transformation and background media in inner sphere. The inconsistent anisotropic acoustic media causes there exist no acoustic  wave solution to satisfy the acostic wave equations and interface continuous  condition global acoustic equations system.  If the pressure wave $P(\vec R)$ is the solution of the acoustic wave equation (14) and satisfy the necessary no scattering interface conditions (26) and (27) on the outer interface  boundary $\vec R = \vec R_2 $, moreover, for infinity countable angular frequencies $\omega _m $ that make $j_1 (k_{b,m} R_1 ) = 0$, then  the  acoustic wave is propagation to penetrate into the inner sphere, $P(\vec R) = \frac{{k_{b,m} ^2 R_1 ^2 }}{{4\pi }}\frac{{e^{ik_{b,m} r{}_s} }}{{\left| {r_s } \right|}}n_1 (k_{b,m} R_1 )j_0 (k_{b,m} R)$,and the wave does satisfy interface continuous condition on the inner interface boundary $\vec R = \vec R_1$, the inner sphere can not be cloaked.  In particular, in this paper, we propose a novel anisotropic media in the sphere $R \le R_1$ that is the induced anisotropic acoustic media  $ (15) \ to \ (17) $ or $ (20) \ to \ (22) $, by the  $0$ to $R_1$ sphere radial transformation (7) or (18),respectively, The anisotropic acoustic media formulas in the sphere $R \le R_1$ is same as that in the annular layer $R_1 \le R \le R_2$. We prove if the anisotropic acoustic media are installed in  the sphere  $R \le R_1$, then the acoustic wave solution of the acoustic wave equation (14) is continuous propagation to penetrate into the inner sphere $R \le R_1$, the inner sphere $R \le R_1$, can not be cloaked. Therefore, 0 to $R_1$ spherical radial transformation can not be used to induce acoustic no scattering cloak.
We propose our Global and Local field method and novel approach to prove the cloak in paper [4] is not "No scattering acoustic cloak". First, we define "the global acoustic equation system" , Second, we difine "acoustic no scattering cloak" as follows that suppose that the anisotropic acoustic media is created in the annular layer that make there exist acoustic wave solution to satisfy the global acoustic equation system, if in the outside of whole sphere the acoustic wave solution equal to incident wave, $p(\vec r) = p_i(\vec r)$, i.e. there exist no scattering wave to disturb the incident wave, and the acoustic wave solution is zero in the inner sphere, i.e. the inner sphere is cloaked, then the annular layer,  $R_1 \le R \le R_2$   and inner sphere $R \le R_1$  is called the "acoustic no scattering cloak. In Statement 7 in this paper, we prove if the induced anisotropic acoustic media (73) - (75) (i.e. (25),(24), (26) in paper [4]) by $0$ to $R_1$ linear transformation "0R1SRLT" in (71)-(22) is installed in the annular layer
$R_1 \le R \le R_2$, and $j_1 (k_b R_1 ) \ne 0$, then there exist no acoustic wave solution
to satisfy the "global acoustic wave eqution system" in (61) to (68). That prove that the cloak in paper [4] is not "acoustic No scattering Cloak".
In the statement 8 in this paper, we prove if the induced anisotropic acoustic media (73) - (75) (i.e. (25),(24),(26) in paper [4]) by $0$ to $R_1$ linear transformation "0R1SRLT" in (71)-(72) is installed in the annular layer
$R_1 \le R \le R_2$, and $j_1 (k_b R_1 ) = 0$, then there exist acoustic wave solution
to satisfy the "global acoustic wave eqution system" in (61) to (68), the nonzero and continuous bounded acoustic wave solution is propagation to penetrate into the inner sphere, the
inner sphere $R \le R_1 $ is not cloaked. That prove that the cloak in paper [4] is not "acoustic No scattering Cloak". 
In the novel statement 9, we prove if the induced anisotropic acoustic media (73) - (75) by $0$ to $R_1$ linear transformation "0R1SRLT" in (71)-(72) is installed in the annular layer
$R_1 \le R \le R_2$ and the inner sphere $R \le R_1$, then there exist acoustic wave solution
to satisfy the "global acoustic wave eqution system" in (61) to (65) and (91) to (93), the nonzero and continuous bounded physical acoustic wave solution is propagation to penetrate into the inner sphere, $R \le R_1 $,the
inner sphere $R \le R_1 $ is not cloaked. That prove that the cloak in paper [4] is not "acoustic No scattering Cloak". Moreover,  that prove that the "Physically ....." explantion in paper [4] is not correct and is a mistake in "No Scattering Cloak Super Physical Sciences". We proved that  0 to $R_1$ spherical radial transformation can not be used to induce acoustic no scattering cloak. The 0 to R1 spherical radial transformation can not be used to induce static electric conductivity no scattering cloak.

Patent are reserved by authors in GL Geophysical Laboratory.
\end{abstract}

\maketitle
\hfill\break \\

\section{\label{sec:level1}INTRODUCTION} 
There are two methods to make electromagnetic invisible cloak, one is our ¡°GILD and GL no scattering modeling and inversion method¡±, another one is transformation method. Using GL no scattering modeling and inversion and without transformation, we proposed electromagnetic (EM) practicable GLLH [1] double layer invisible cloak with relative refractive index not less than 1 and EM GLHUA[2] double layer invisible cloak with relative parameters not less than 1, both without exceeding light speed propagation. In paper [2], we discover a novel electromagnetic invisible cloak media with relative parameter not less than 1 in the sphere layer $R_1 \le R \le R_2$. In particular, we find exact analytic electromagnetic wave propagation in the novel practicable double layer electromagnetic invisible cloak in the paper [2].  Using 0 to $R_1$ sphere radial linear transformation, Pendry et al proposed EM invisible cloak [3]. It is proved that by [5][6][7][8] the Pendry EM cloak [3] is invisible cloak with infinite speed and exceeding light phase velocity  fundamental difficulties.
In this paper, we define "the acoustic no scattering cloak" as follows, given $R_2 > R_1 > 0$, a whole sphere $R \le R_2$ is located in the background acoustic space, to divide the sphere $R \le R_2$ into an annular layer $R_1 \le R \le R_2$ and an inner sphere $R \le R_1$, the 3D full Space is split into the three domains and two interfaces $\vec R = \vec R_2$ and $\vec R = \vec R_1$, the outside of the whole sphere $R \ge R_2$ with background isotropic acoustic media, the annular layer $R_1 \le R \le R_2$ with anisotropic acoustic media, the inner sphere $R \le R_1$ with background isotropic acoustic media, the 3 acoustic equations in the three domains  respectively  and the 4 interface continuous conditions equations of wave and its derivative on the interfaces $\vec R = \vec R_2$ and $\vec R = \vec R_1$ that compose global acoustic equation system. Suppose that the anisotropic acoustic media is created in the annular layer that make there exist acoustic wave solution to satisfy the global acoustic equation system with Sommerfeld far field radiation condition, if in the outside of whole sphere the acoustic wave solution equal to incident wave, i.e. there exist no scattering wave to disturb the incident wave, and the acoustic wave solution is zero in the inner sphere, i.e. the inner sphere is cloaked, then the annular layer and inner sphere is called the acoustic no scattering cloak, the anisotropic acoustic media is called no scattering materials, the annular layer is called no scattering cloaking layer, the inner sphere is called no scattering cloaked concealment.  
We prove that the acoustic cloak in paper [4] is not "no scattering acoustic cloak". We prove that the  $0$ to $R_1$  spherical radial transformation method can not be used to induce acoustic no scattering cloak, and prove that  the induce anisotropic acoustic media in annular layer $R_1 \le R \le R_2$ by linear $0$ to $R_1$ sphere radial transformation in [4] is inconsistent with background isotropic media in inner sphere that causes there exist no acoustic wave solution to satisfy the above global acoustic equation system.
Because in the $0$ to 
$R_1$ radial spherical coordinate transformation, the value range of the acoustic field is invariant, Let $p_i(\vec r)$ be incident pressure acoustic wave in background space before transformation, $\vec r$ be radial variable in background space before transformation, $\vec R$ be radial variable after transformation. Let the $0$ to $R_1$ sphere radial transformation to be $R(r) = R_1  + Q(r)$, and inverse transformation to be $r = Q^{ - 1} (R - R_1)$, the $P_1 (\vec R)$ be pressure acoustic wave in the sphere annular layer $R_1 \le R \le R_2$ after transformation,  $P_1 (R) = p_i(r(R)) = p_i(Q^{ - 1} (R - R_1 ))$, $p(\vec r)$ be pressure acoustic wave in background space outside sphere $r \ge R_2$, if $p(\vec r)=p_i(\vec r)$
then there exist no scattering wave to disturb the incident wave, the $P_2 (\vec R)$ be pressure acoustic wave in the inner sphere $R \le R_1$, Suppose that on the outer interface boundary $\vec R =\vec R_2 $, acoustic wave is continuous and its radial derivative with parameter is also continuous between acoustic wave $P_1(\vec R)$ and incident acoustic wave $p_i(\vec r)$ , $P_1(\vec {R_2}^-)=p_i(\vec {R_2}^+)$ and $\frac{1}{{\rho _r }}R_2^2 \frac{\partial }{{\partial R}}P(\vec R_2 ^ -  ) = R_2^2 \frac{\partial }{{\partial r}}p_i (\vec R_2 ^ +  )$,   then there exist no scattering acoustic wave to disturb the incident wave $p_i(\vec r)$ in outside sphere, both interface continuous conditions are necessary condition for "there exist no scattering acoustic wave from whole sphere". We call the above acoustic wave $P_1(\vec R)$ to be "NO Scattering Acoustic Wave" that does satisfy "no scattering acoustic wave condition" (26) and (27) on the outer interface boundary $\vec R = \vec R_2$. If the acoustic wave satisfies the no scattering interface conditions (26) and (27) on the outer interface boundary $\vec R = \vec R_2$,  then on the  inner interface boundary $\vec R = \vec R_1$, the acoustic wave $P_1(\vec R)$ is nonzero constant $P_1 (R_1 ^ +  ) \ne 0$ , and then the inner sphere $R \le R_1 $ can not be cloaked, otherwise, the ¡°$P_2(\vec R) = 0$ in $R \le R_1 $ ¡±that  causes contradiction between $P_2(R_1 ^ -  ) = 0$ and $P_1(R_1 ^ +  ) \ne 0$,$P_2(R_1 ^ -  ) \ne P_1(R_1 ^ +  ) $, the basic interface continuous conditions are destroyed,  Inversely, if the inner sphere is cloaked that will cause the pressure wave does not satisfy no scattering interface condition on the outer interface boundary  $\vec R= \vec R_2 $,
$P(R_2 ^ -  ) \ne p_i (R_2 ^ +  )$. In particular, in this paper,we propose a novel anisotropic media in the sphere $R \le R_1$ that is the induced anisotropic acoustic media  $ (15) \ to \ (17) $ or $ (20) \ to \ (22) $, by the  $0$ to $R_1$ sphere radial transformation (7) or(18),respectively, The  anisotropic acoustic media formulas in the sphere $R \le R_1$ is same as that in the annular layer $R_1 \le R \le R_2$. We prove if the anisotropic acoustic media are installed in  the sphere  $R \le R_1$, then the acoustic wave solution of the acoustic wave equation (14) is continuous propagation to penetrate into the inner sphere $R \le R_1$, the inner sphere $R \le R_1$, can not be cloaked. Therefore, 0 to $R_1$ spherical radial transformation can not be used to induce acoustic no scattering cloak.
  In some published papers on acoustic cloak, the $0$ to $R_1$ spherical radial transformation was wrong used to induce their acoustic cloak, we prove  the cloak in paper [4] that is not "no scattering acoustic cloak". 
We propose our Global and Local field method and novel approach to prove the cloak in paper [4] is not "No scattering acoustic cloak". First, we define "the global acoustic equation system" , Second, we difine "acoustic no scattering cloak" as follows that suppose that the anisotropic acoustic media is created in the annular layer that make there exist acoustic wave solution to satisfy the global acoustic equation system, if in the outside of whole sphere the acoustic wave solution equal to incident wave, $p(\vec r) = p_i(\vec r)$, i.e. there exist no scattering wave to disturb the incident wave, and the acoustic wave solution is zero in the inner sphere, i.e. the inner sphere is cloaked, then the annular layer,  $R_1 \le R \le R_2$   and inner sphere $R \le R_1$  is called the "acoustic no scattering cloak. In Statement 7 in this paper, we prove if the induced anisotropic acoustic media (73) - (75) (i.e. (25),(24), (26) in paper [4]) by $0$ to $R_1$ linear transformation "0R1SRLT" in (71)-(22) is installed in the annular layer
$R_1 \le R \le R_2$, and $j_1 (k_b R_1 ) \ne 0$, then there exist no acoustic wave solution
to satisfy the "global acoustic wave eqution system" in (61) to (68). That prove that the cloak in paper [4] is not "No scattering Cloak".
In the statement 8 in this paper, we prove if the induced anisotropic acoustic media (73) - (75) (i.e. (25),(24),(26) in paper [4]) by $0$ to $R_1$ linear transformation "0R1SRLT" in (71)-(72) is installed in the annular layer
$R_1 \le R \le R_2$, and $j_1 (k_b R_1 ) = 0$, then there exist acoustic wave solution
to satisfy the "global acoustic wave eqution system" in (61) to (68), the nonzero and continuous bounded acoustic wave solution is propagation to penetrate into the inner sphere, the
inner sphere $R \le R_1 $ is not cloaked. That prove that the cloak in paper [4] is not "No scattering Cloak". 
In the novel statement 9, we prove if the induced anisotropic acoustic media (73) - (75) by $0$ to $R_1$ linear transformation "0R1SRLT" in (71)-(72) is installed in the annular layer
$R_1 \le R \le R_2$ and the inner sphere $R \le R_1$, then there exist acoustic wave solution
to satisfy the "global acoustic wave eqution system" in (61) to (65) and (91) to (93), the nonzero and continuous bounded physical acoustic wave solution is propagation to penetrate into the inner sphere, $R \le R_1 $,the
inner sphere $R \le R_1 $ is not cloaked. That prove that the cloak in paper [4] is not "No scattering Cloak". Moreover,  that prove that the "Physically ....." explantion in paper [4] is not correct and is a mistake in "No Scattering Cloak Super Physical Sciences". We proved that  0 to $R_1$ spherical radial transformation can not be used to induce acoustic no scattering cloak. The 0 to R1 spherical radial transformation can not be used to induce static electric conductivity no scattering cloak.
The contents of this paper are as follows. The isotropic pressure acoustic wave equation in the 3D spherical coordinate  space is described in section 2. In section 3, we describe that $0$ to $R_1$  spherical radial coordinate transformation does induce relative acoustic media for pressure acoustic wave propagation. In section 4, we prove that if acoustic wave solution does satisfy necessary no scattering conditions on the outer interface boundary $\vec R = \vec R_2 $, then the pressure wave is nonzero constant on the inner interface  boundary. In section 5,
we proved that cloaked inner sphere that causes the wave solution of the acoustic equation (14) is different from the incident wave on outer interface boundary, $\vec R = \vec R_2 $ 
In section 6, we prove that the $0$ to $R_1$  spherical radial transformation method can not be used to induce acoustic no scattering acoustic cloak, in this section
we proposal a novel anisotropic acoustc media in the inner sphere that making
the acoustic wave solution of the acoustic equation (14) can be continuous propagation to penetrate into the inner sphere $R \le R_1$, the inner sphere can be not cloaked. The discussions is described in section 7, We write physical letter to summary this paper in section 8, in this section, we propose our Global and Local field method and novel approch to prove the cloak in paper [4] is not "No scattering acoustic cloak". Conclusion is presented in section 9.
\begin{figure}[b]
\centering
\includegraphics[width=0.86\linewidth,draft=false]{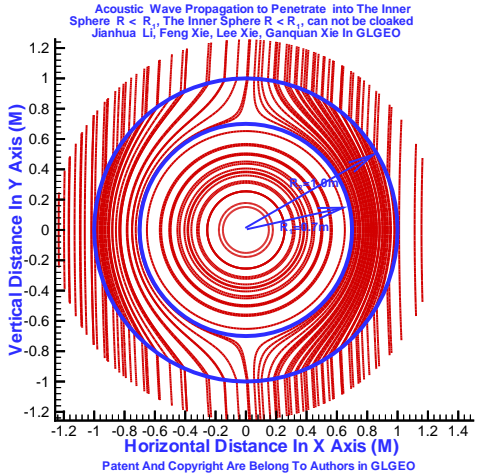}
\caption{ (color online) 
Pressure acoustic wave propagation through the annular layer
$R_1 \le R \le R_2$ and inner sphere $R \le R_1$,the anisotropic
acoustic media $(20) \ - \ (22)$ are installed in the annular layer
and the inner sphere, the scattering acoustic wave is propagation
to penetrate into the inner sphere $R \le R_1$,the inner sphere is not
"acoustic no scattering acoustic cloak", where $R_1=0.7m$,$R_2=1.0m$}\label{fig1}
\end{figure}
\begin{figure}[b]
\centering
\includegraphics[width=0.86\linewidth,draft=false]{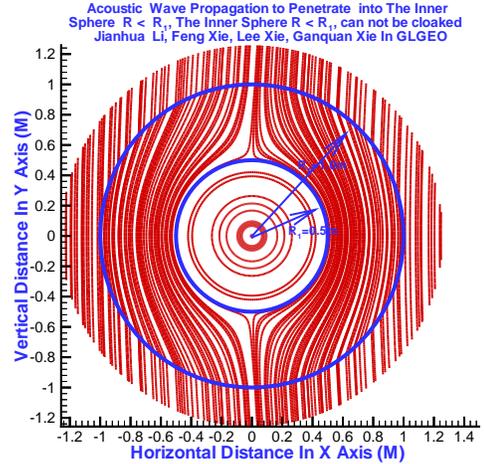}
\caption{ (color online) 
Pressure acoustic wave propagation through the annular layer
$R_1 \le R \le R_2$ and inner sphere $R \le R_1$,the anisotropic
acoustic media $(20) \ - \ (22)$ are installed in the annular layer
and the inner sphere, the scattering acoustic wave is propagation
to penetrate into the inner sphere $R \le R_1$,the inner sphere is not
"acoustic no scattering acoustic cloak", where $R_1=0.5m$,$R_2=1.0m$}\label{fig2}
\end{figure}
\section {The isotropic acoustic wave equation in the 3D spherical coordinate space}
\subsection {The isotropic pressure acoustic wave equation in the  3D  spherical coordinate space with background acoustic speed  and incident pressure acoustic wave
}
In the 3D spherical coordinate in  space, the isotropic pressure acoustic wave equation 
is Helmholtz equation as follows:
\begin{equation}
\begin{array}{l}
 \frac{\partial }{{\partial r}}\left( {r^2 \frac{{\partial p(\vec r)}}{{\partial r}}} \right) + \frac{1}{{\sin \theta }}\frac{\partial }{{\partial \theta }}\sin \theta \frac{{\partial p(\vec r)}}{{\partial \theta }} \\ 
  + \frac{1}{{\sin ^2 \theta }}\frac{{\partial ^2 p(\vec r)}}{{\partial \phi ^2 }} + k_b^2 r^2 p(\vec r) = S(\vec r,\vec r_s ), \\ 
 \end{array}
\end {equation}                
where $p(\vec r)$ is the pressure wave field in the background space. $c_b$ is the background constant acoustic speed, for example, the pressure wave in background deep ocean, $\omega $ is the angular frequency, let $k_b  = \frac{\omega }{{c_b }}$, $S(\vec r,\vec r_s )$ is the acoustic source located in $\vec r_s $ in some device in ocean.  For example,     

\begin {equation}
      S(\vec r,\vec r_s ) = \frac{1}{{\sin \theta }}\delta (r - r_s )\delta (\theta  - \theta _s )\delta (\phi  - \phi _s ),
\end {equation}
The incident pressure wave field $p_i (\vec r)$ does satisfy the acoustic equation (1) in infinite space which is excited by source (2),.
\begin {equation}
p_i (\vec r) =  - \frac{1}{{4\pi }}\frac{{e^{ik_b \left| {\vec r - \vec r_s } \right|} }}{{\left| {\vec r - \vec r_s } \right|}},
\end {equation}
\subsection {"Background no scattering sphere" $r \le R_2 $ and homogeneous acoustic wave equation}
Suppose that the source is located outside the whole sphere with radius $R_2  > 0$. $r_s  > R_2 $,  in the sphere $r \le R_2 $ ,with background acoustic medium , the pressure acoustic wave equation (1) become to the homogeneous acoustic wave equation,
\begin{equation}
\begin{array}{l}
 \frac{\partial }{{\partial r}}\left( {r^2 \frac{{\partial p(\vec r)}}{{\partial r}}} \right) + \frac{1}{{\sin \theta }}\frac{\partial }{{\partial \theta }}\sin \theta \frac{{\partial p(\vec r)}}{{\partial \theta }} \\ 
  + \frac{1}{{\sin ^2 \theta }}\frac{{\partial ^2 p(\vec r)}}{{\partial \phi ^2 }} + k_b ^2 r^2 p(\vec r) = 0, \\ 
 \end{array}
\end{equation}
on the interface boundary $\vec r = \vec R_2 $, the pressure acoustic wave field $p(\vec r)$ and its radial derivative satisfy the continuous interface conditions
\begin{equation}
p(\vec R_2 ^ -  ) = p(\vec R_2 ^ +  ) = p_i (\vec R_2 ^ +  )
\end{equation}
\begin {equation}
\frac{\partial }{{\partial r}}p(\vec R_2 ^ -  ) = \frac{\partial }{{\partial r}}p(\vec R_2 ^ +  ) = \frac{\partial }{{\partial r}}p_i (\vec R_2 ^ +  )
\end{equation}                                                 
where $p_i (\vec r)$ is incident acoustic wave in (3), $\vec R_2 ^ +   = (R_2 ^ +  ,\theta ,\phi )$.$\vec R_2 ^ -   = (R_2 ^ -  ,\theta ,\phi )$
The $p(\vec r)$ in (3) in sphere $r \le R_2 $ is the solution of equation (4) with continuous interface conditions (5) and (6). The sphere $r \le R_2 $ with background acoustic medium is called the ¡°background no scattering sphere¡±, the interface continuous conditions (5) and (6) on the interface boundary $r = R_2$  is sufficient and necessary conditions for no scattering from the whole sphere $r \le R_2 $  to disturb the incident wave in outside of whole sphere in $r \ge R_2 $
\section {$0$ to $R_1$  spherical radial coordinate transformation does induce relative acoustic media for pressure acoustic wave propagation}
\subsection{$0$ to $R_1$ sphere radial coordinate transformation}
For $R_1  > 0$, $R_2 >R_1$, inside of the sphere $r \le R_2 $
, the $0$ to $R_1$ sphere radial continuous coordinate transformation is
\begin {equation}
R(r) = R_1  + Q(r),0 \le r \le R_2,
\end {equation}
\begin {equation}
R(0) = R_1 ,Q(0) = 0,
\end {equation}
\begin {equation}
 R(R_2 ) = R_2,Q(R_2 ) = R_2  - R_1 , 
\end {equation}
\begin {equation}
\frac{\partial }{{\partial r}}R(r) \ge 0,
\end {equation}

where $R = R(r)$ is radial coordinate in the physical spherical annular layer $R_1  \le R \le R_2 $.
We use $r$,$\vec r$ ,$p(\vec r)$ to denote sphere radial variable, sphere radial vector, and pressure acoustic wave propagation in background space, and use $R$,$\vec R$,$P(\vec R)$ to denote sphere radial variable, sphere radial vector, and pressure acoustic wave propagation in sphere
annular layer space domain, $R_1  \le R \le R_2 $, respectively.

By $0$ to $R_1$ sphere radial transformation (7)-(10), the "background no scattering background sphere"  is compressed into the physical sphere annular layer domain, $R_1 \le R \le R_2$ and a new absolute empty topology 3D inner sphere $R \le R_1$ is created. We will install some  acoustic media in the inner sphere, for example, the background medium is in the inner sphere, $R \le R_1$, We prove that the background medium  in the inner sphere is inconsistent with the induced anisotropic acoustic media by transformation in the annular layer. 

The inverse transformation of (7) is
\begin{equation}
r = r(R) = Q^{ - 1} \left( {R - R_1 } \right),R_1  \le R \le R_2 ,
\end {equation}

\subsection {The induced acoustic equation with relative anisotropic acoustic media}
 
To substitute $0$ to $R_1$   radial coordinate transformation (7)-(10) into  the pressure acoustic equation (4) in the background sphere $r \le R_2 $, and by the following transformations,
Let  $\vec R = (R,\theta ,\phi ) = (R(r),\theta ,\phi )$
\begin{equation}
\begin{array}{l}
 \frac{{\partial R}}{{\partial r}}\frac{\partial }{{\partial R}}\left( {R^2 \frac{{r^2 }}{{R^2 }}\frac{{\partial R}}{{\partial r}}\frac{{\partial p(\vec r)}}{{\partial R}}} \right) + \frac{1}{{\sin \theta }}\frac{\partial }{{\partial \theta }}\sin \theta \frac{{\partial p(\vec r)}}{{\partial \theta }} \\ 
  + \frac{1}{{\sin ^2 \theta }}\frac{{\partial ^2 p(\vec r)}}{{\partial \phi ^2 }} + k_b ^2 \frac{{r^2 }}{{R^2 }}R^2 p(\vec r) = 0, \\ 
 \end{array}
\end {equation}
\begin{equation}
\begin{array}{l}
 \frac{\partial }{{\partial R}}\left( {\frac{{R^2 }}{{\frac{{R^2 }}{{r^2 }}\frac{{\partial r}}{{\partial R}}}}\frac{{\partial p(\vec r)}}{{\partial R}}} \right) + \frac{1}{{\frac{{\partial R}}{{\partial r}}}}\frac{1}{{\sin \theta }}\frac{\partial }{{\partial \theta }}\sin \theta \frac{{\partial p(\vec r)}}{{\partial \theta }} \\ 
  + \frac{1}{{\frac{{\partial R}}{{\partial r}}}}\frac{1}{{\sin ^2 \theta }}\frac{{\partial ^2 p(\vec r)}}{{\partial \phi ^2 }} + k_b ^2 \frac{1}{{\frac{{R^2 }}{{r^2 }}\frac{{\partial R}}{{\partial r}}}}R^2 p(\vec r) = 0, \\ 
 \end{array}
\end{equation}
by substitution $P_1 (R) = p_i(r(R)) = p_i(Q^{ - 1} (R - R_1 ))$, 
the equation (4) is translated to the following anisotropic homogeneous acoustic equation in physical sphere annular layer domain, $R_1  \le R \le R_2 $
\begin{equation}
\begin{array}{l}
 \frac{\partial }{{\partial R}}\frac{1}{{\rho _r }}R^2 \frac{{\partial P_1(\vec R)}}{{\partial R}} + \frac{1}{{\rho _\theta  \sin \theta }}\frac{\partial }{{\partial \theta }}\sin \theta \frac{{\partial P_1 (\vec R)}}{{\partial \theta }} \\ 
  + \frac{1}{{\rho _\phi  \sin ^2 \theta }}\frac{{\partial ^2 P_1 (\vec R)}}{{\partial \phi ^2 }} + \frac{1}{{K_t }}k_b ^2 R^2 P_1(\vec R) = 0, \\ 
 \end{array}
\end{equation}   

by the $0$ to $R_1$ sphere radial transformation (7-10), the induced relative anisotropic parameters in physical sphere annular layer domain $R_1  \le R \le R_2 $ are

\begin{equation}
\rho _r  = \frac{{R^2 }}{{r^2 }}\frac{{dr}}{{dR}},
\end{equation}
\begin{equation}
\rho _\theta   = \rho _\phi   = \frac{{dR}}{{dr}},
\end{equation}
\begin{equation}
K_t  = \frac{{R^2 }}{{r^2 }}\frac{{dR}}{{dr}}.
\end{equation}
Use $P(\vec R)$ to replace $P_1(\vec R)$ in the equation (14), the equation (14) is the general anisotropic acoustic equation.
In particular, for $0$ to $R_1$ sphere radial linear transformation in the paper [4],
\begin{equation}
R = R_1  + Q(r) = R_1  + \frac{{R_2  - R_1 }}{{R_2 }}r,
\end{equation}
the linear transformation (18) does satisfy transformation conditions (7)-(10), and its inverse  transformation
\begin{equation}
r = Q^{ - 1} (R - R_1 ) = \frac{{R_2 }}{{R_2  - R_1 }}(R - R_1 ),
\end{equation}
The anisotropic relative parameters induced by linear transformation (18) are
\begin{equation}
\rho _r  = \frac{{R^2 }}{{r^2 }}\frac{{dr}}{{dR}} = \frac{{(R_2  - R_1 )}}{{R_2^{} }}\frac{{R^2 }}{{(R - R_1 )^2 }},
\end{equation}
\begin{equation}
\rho _\theta   = \rho _\phi   = \frac{{dR}}{{dr}} = \frac{{R_2  - R_1 }}{{R_2 }},
\end{equation}
\begin{equation}
K_t  = \frac{{R^2 }}{{r^2 }}\frac{{dR}}{{dr}} = \frac{{(R_2  - R_1 )^3 }}{{R_2^3 }}\frac{{R^2 }}{{(R - R_1 )^2 }},
\end{equation}

The inner radius is $a$ and outer radius is $b$ the paper [4], corresponding, inner radius is  $R_1$ ,outer radius is $R_2$ in this paper, $a$ is $R_1$ ,$b$ is $R_2$, respectively,  by $0$ to $R_1$ sphere radial linear transformation (18), relative anisotropic acoustic parameters in $(20)\ to \ (22)$ in this paper is same as the relative acoustic parameters [(24)-(26)] in paper [4] induced by same linear transformation in (18).

Substitute the background acoustic medium,
\begin{equation}
\rho _r  = \rho _\theta   = \rho _\phi   = K_t  = 1,
\end{equation}
into the equation (14), the anisotropic pressure acoustic equation (14) becomes to isotropic homogeneous pressure acoustic equation (4) without source. On the interface boundary $\vec r = \vec R = \vec R_2$, the general pressure wave field solution of equation (14), $P(\vec R)$, and pressure acoustic wave solution of equation (1), $p(\vec r)$, satisfy the following continuous interface conditions
\begin {equation}
P(\vec R_2 ^ -  ) = p(\vec R_2 ^ +  ),
\end{equation}
and their derivatives satisfy the following continuous interface conditions
\begin{equation}
\frac{1}{{\rho _r }}R_2^2 \frac{\partial }{{\partial R}}P(\vec R_2 ^ -  ) = R_2^2 \frac{\partial }{{\partial r}}p(\vec R_2 ^ +  ),
\end{equation}
where $\vec R_2 ^ -   = (R_2 ^ -  ,\theta ,\phi )$, $\vec R_2 ^ +   = (R_2 ^ +  ,\theta ,\phi )$. If substitute the pressure acoustic wave $P_1 (\vec R)$ by transformation of $p_i(\vec r)$ and  the incident wave and its derivative into right hand of (24) and (25), the interface continuous conditions (24) and (25) become no scattering interfac
e continuous conditions  (26)  and (27),
\begin{equation}
P_1(\vec R_2 ^ -  ) = p_i (\vec R_2 ^ +  ),
\end{equation}
\begin{equation}
\frac{1}{{\rho _r }}R_2^2 \frac{\partial }{{\partial R}}P_1(\vec R_2 ^ -  ) = R_2^2 \frac{\partial }{{\partial r}}p_i (\vec R_2 ^ +  ),
\end {equation}
equation (26) and (27)  are the sufficient and necessary condition of that no scattering acoustic wave
from the whole sphere $R \le R_2 $ to disturb incident wave in outside of the whole sphere. The pressure acoustic wave $P_1(\vec R)$ is no scattering acoustic wave by transformation of the incident wave $p_i(\vec r)$.

\section {Outer interface no scattering continuous conditions govern that the solution of acoustic wave equation is nonzero constant on the inner interface boundary}

Outer interface continuos no scattering conditions  (26) and (27) govern that the pressure acoustic wave solution of acoustic equation (14) is nonzero constant on the inner interface boundary $\vec r = \vec R_1$, 
$P_1(\vec R_1 ) =  - \frac{1}{{4\pi }}\frac{{e^{ik_b \left| {r_s } \right|} }}{{\left| {r_s } \right|}} \ne 0.$

${\bold Statement \ 1:}$  Suppose that the point delta source (2) is in outside of whole sphere, $r_s  > R_2 $, the pressure wave $P_1(\vec R)$ does satisfy the acoustic equation (14) with relative anisotropic parameters (15)-(17) or (20)-(22) by transformation (7) or (18), respectively, and does satisfy the wave and derivative no scattering continuous interface continuous conditions (26) and (27) , for the transformation (7) and induced anisotropic parameters (15)-(17), then $P_1(\vec R)$ has the analytic express
\begin{equation}
\begin{array}{l}
 P_1(\vec R) = p_i(r(R),\theta ,\phi ) \\ 
  =  - \frac{1}{{4\pi }}\frac{{e^{ik_b \left| {(Q^{ - 1} (R - R_1 ),\theta ,\phi ) - \vec r_s } \right|} }}{{\left| {(Q^{ - 1} (R - R_1 ),\theta ,\phi ) - \vec r_s } \right|}}, \\ 
 \end{array}
\end{equation}
Where $r = Q^{ - 1} (R - R_1 )$.  For $P_1(\vec R)$ is solution of he acoustic equation (14) with relative anisotropic parameters (20)-(22) by linear transformation (18)  
\begin{equation}
\begin{array}{l}
 P_1(\vec R) = p_i(r(R),\theta ,\phi ) \\ 
  =  - \frac{1}{{4\pi }}\frac{{e^{ik_b \left| {(Q^{ - 1} (R - R_1 ),\theta ,\phi ) - \vec r_s } \right|} }}{{\left| {(Q^{ - 1} (R - R_1 ),\theta ,\phi ) - \vec r_s } \right|}} \\ 
  =  - \frac{1}{{4\pi }}\frac{{e^{ik_b \left| {(\frac{{R_2 }}{{R_2  - R_1 }}(R - R_1 ),\theta ,\phi ) - \vec r_s } \right|} }}{{\left| {(\frac{{R_2 }}{{R_2  - R_1 }}(R - R_1 ),\theta ,\phi ) - \vec r_s } \right|}}, \\ 
 \end{array}
\end{equation}
${\bold Proof:}$  By inverse transformation in (11) or (19), $P_1 (\vec R)$ is put back to $p_i(\vec r)$  
\begin{equation}
\begin{array}{l}
 left \ hand \ side \ of \  (14) \\ 
  = \frac{{\partial r}}{{\partial R}}\frac{\partial }{{\partial r}}\frac{1}{{\rho _r }}R^2 \frac{{\partial r}}{{\partial R}}\frac{{\partial p_i(\vec r)}}{{\partial r}} + \frac{1}{{\rho _\theta  \sin \theta }}\frac{\partial }{{\partial \theta }}\sin \theta \frac{{\partial p_i(\vec r)}}{{\partial \theta }} \\ 
  + \frac{1}{{\rho _\phi  \sin ^2 \theta }}\frac{{\partial ^2 p_i(\vec r)}}{{\partial \phi ^2 }} + \frac{1}{{K_t }}k_b ^2 R^2 p_i(\vec r), \\ 
 \end{array}
\end{equation}
Substitute $\rho _r  = \frac{{R^2 }}{{r^2 }}\frac{{dr}}{{dR}}$ in (15), $\rho _\theta   = \rho _\phi   = \frac{{dR}}{{dr}}$ in (16) and $K_t  = \frac{{R^2 }}{{r^2 }}\frac{{dR}}{{dr}}$ in (17) into the equation (30), we have
\begin{equation}
\begin{array}{l}
 left \ hand \ side \ of \ (14) \\ 
  = \frac{\partial }{{\partial r}}\left( {r^2 \frac{{\partial p_i(\vec r)}}{{\partial r}}} \right) + \frac{1}{{\sin \theta }}\frac{\partial }{{\partial \theta }}\sin \theta \frac{{\partial p_i(\vec r)}}{{\partial \theta }} \\ 
  + \frac{1}{{\sin ^2 \theta }}\frac{{\partial ^2 p_i(\vec r)}}{{\partial \phi ^2 }} + k_b ^2 r^2 p_i(\vec r) = 0. \\ 
 \end{array}
\end{equation}

 the right hand side of (31) is the equation (4), because $p_i(\vec r)$ satisfies equation (4), therefore, $P_1(\vec R)$ in (28) or (29) does satisfy the acoustic equation (14). Because the interface boundary $\vec R = \vec R_2$ is invariant under transformation condition (7) or (18), and by interface condition (5), we have
\begin{equation}
P_1(\vec R_2 ^ -  ) = p_i(\vec R_2 ^ -  ) = p_i (\vec R_2 ^ +  ),
\end{equation}

$P_1(\vec R)$ in (28) does satisfy the no scattering interface condition (27).
\begin{equation}
\begin{array}{l}
 left \ hand\  side \ of \ (27) \\ 
  = \frac{1}{{\rho _r }}R_2^2 \frac{\partial }{{\partial R}}P_1(\vec R_2 ^ -  ) \\ 
  = \frac{1}{{\rho _r }}R_2^2 \frac{{\partial r}}{{\partial R}}\frac{\partial }{{\partial r}}P_1(\vec R_2 ^ -  ), \\ 
 \end{array}
\end{equation}
Substitute $\rho _r  = \frac{{R^2 }}{{r^2 }}\frac{{dr}}{{dR}}$ in (15) into (33), by transformation condition (9) and derivative interface continuous continuous condition (6), we have 
\begin{equation}
\begin{array}{l}
 left \ hand \ side \ of \ (27) \\ 
  = \frac{1}{{\rho _r }}R_2^2 \frac{\partial }{{\partial R}}P_1(\vec R_2 ^ -  ) \\ 
  = \frac{{r^2 }}{{R_2 ^2 }}|_{r = R_2 } R_2 ^2 \frac{\partial }{{\partial r}}P_1(\vec R_2 ^ -  ) = R_2 ^2 \frac{\partial }{{\partial r}}p_i(\vec R_2 ^ +  ), \\ 
 \end{array}
\end{equation}
the derivative interface continuous  condition (27) is satisfied . Substitute 
$r = Q^{ - 1} (R - R_1 ) = \frac{{R_2 }}{{R_2  - R_1 }}(R - R_1 )
$ in (19) into (28), we obtain (29). The Statement 1  is proved.

${\bold Statement \ 2:}$  Suppose that the point delta source (2) is in outside of whole sphere, $r_s  > R_2$,  the pressure acoustic wave $P_1(\vec R)$ is solution of the pressure acoustic equation (14) in the sphere annular layer domain $R_1  \le R \le R_2 $ with anisotropic relative acoustic parameters in (15)-(17) or (20)-(22), also, the pressure acoustic wave $P_1(\vec R)$  satisfies no scattering interface continuous conditions (26) and (27) on the interface boundary $\vec R = \vec R_2 $, then on the inner interface boundary $\vec R = \vec R_1 $,  $P_1(\vec R_1 )$ is nonzero constant, 
\begin{equation}
P_1(\vec R_1 ) =  - \frac{1}{{4\pi }}\frac{{e^{ik_b \left| {r_s } \right|} }}{{\left| {r_s } \right|}} \ne 0.
\end{equation}
Proof:  In this Statement , outer interface boundary is $\vec R = \vec R_2  = \left( {R_2 ,\theta ,\phi } \right)$,
Inner interface boundary is $\vec R = \vec R_1  = \left( {R_1 ,\theta ,\phi } \right)$. Because pressure acoustic wave $P_1(\vec R)$ is solution of the pressure acoustic equation (14) in the sphere annular layer domain $R_1  \le r \le R_2 $ with anisotropic  relative acoustic parameters in (15)-(17) by transformation (7), and also $P_1(\vec R)$ does satisfy the no scattering interface conditions (26)-(27), by (28) in Statement 1,
\begin{equation}
P_1(\vec R) =  - \frac{1}{{4\pi }}\frac{{e^{ik_b \left| {(Q^{ - 1} (R - R_1 ),\theta ,\phi ) - \vec r_s } \right|} }}{{\left| {(Q^{ - 1} (R - R_1 ),\theta ,\phi ) - \vec r_s } \right|}},
\end{equation}
then on the inner interface boundary $\vec R=\vec R_1$, by transform property (8), we have
\begin{equation}
\begin{array}{l}
 P_1(\vec R_1 ) =  - \frac{1}{{4\pi }}\frac{{e^{ik_b \left| {(Q^{ - 1} (R_1  - R_1 ),\theta ,\phi ) - \vec r_s } \right|} }}{{\left| {(Q^{ - 1} (R_1  - R_1 ),\theta ,\phi ) - \vec r_s } \right|}} \\ 
  =  - \frac{1}{{4\pi }}\frac{{e^{ik_b \left| {(0,\theta ,\phi ) - \vec r_s } \right|} }}{{\left| {(0,\theta ,\phi ) - \vec r_s } \right|}} =  - \frac{1}{{4\pi }}\frac{{e^{ik_b \left| {r_s } \right|} }}{{\left| {r_s } \right|}} \ne 0. \\ 
 \end{array}
\end{equation}

For $P_1(\vec R) $ is solution of the pressure acoustic equation (14) in the sphere annular layer domain $R_1  \le r \le R_2 $ with anisotropic relative acoustic parameters in (20)-(22) by linear transformation (18), Because pressure acoustic wave $P_1(\vec R)$ also satisfies the no scattering interface continuous conditions (26)-(27), by (29) in Statement 1,
\begin{equation}
\begin{array}{l}
 P_1(\vec R_1 ) =  - \frac{1}{{4\pi }}\frac{{e^{ik_b \left| {(\frac{{R_2 }}{{R_2  - R_1 }}(R - R_1 ),\theta ,\phi ) - \vec r_s } \right|} }}{{\left| {(\frac{{R_2 }}{{R_2  - R_1 }}(R - R_1 ),\theta ,\phi ) - \vec r_s } \right|}}|_{R = R_1 }  \\ 
  =  - \frac{1}{{4\pi }}\frac{{e^{ik_b \left| {(0,\theta ,\phi ) - \vec r_s } \right|} }}{{\left| {(0,\theta ,\phi ) - \vec r_s } \right|}} =  - \frac{1}{{4\pi }}\frac{{e^{ik_b \left| {r_s } \right|} }}{{\left| {r_s } \right|}} \ne 0. \\ 
 \end{array}
\end{equation}

The Statement 2 is proved.

\section { Cloaked inner sphere that causes the wave solution of the acoustic equation (14) is different from the incident wave on outer interface boundary, $\vec R = \vec R_2 $.}

In this section, we propose if inner sphere is cloaked,  then the wave solution of the anisotropic acoustic equation (14)  is different from the incident wave on outer interface boundary, $\vec R = \vec R_2 $.  By $0$ to $R_1$  sphere radial transformation (7)-(11) or (18)-(19), the background sphere $R \le R_2$ is compressed into the physical sphere annular layer domain, $R_1  \le R \le R_2$. A new sphere $R \le R_1$ is inflated from ¡°zero¡±. Originally, the new sphere $R \le R_1$ is absolute empty topological space without physical acoustic geometry medium. We should make acoustic media in the new inner sphere $R \le R_1$ . We prove the  background acoustic speed medium $c_b$  in the inner sphere $R \le R_1$  that is inconsistent with the induced anisotropic acoustic media in  $R_1  \le R \le R_2$ by $0$ to $R_1$  sphere radial transformation. 

Suppose that the inner sphere $R \le R_1$ is cloaked, pressure acoustic wave $P_2(\vec R) = 0$ in inner sphere, it obvious $P_2(\vec R) = 0$  is solution of acoustic equation (4) in  inner sphere $R \le R_1$  with background medium. Let $P(\vec R)$ is wave solution of the anisotropic acoustic equation (14) in the annular layer, $R_1  \le R \le R_2$. The  $P_2(\vec R)=0$ in $R \le R_1$  induces
\begin{equation}
P(\vec R^ +  _1 ) = P_2(\vec R^ -  _1 ) = 0,
\end{equation}
\begin{equation}
\begin{array}{l}
 \frac{1}{{\rho _r }}R_1 ^2 \frac{\partial }{{\partial R}}P(\vec R_1 ^ +  ) =  \\ 
  = (Q^{ - 1} (R - R_1 ))^2 \frac{{dR}}{{dr}}\frac{\partial }{{\partial R}}P(\vec R_1 ^ +  ) \\ 
  = R_1 ^2 \frac{\partial }{{\partial r}}P_2(\vec R_1 ^ -  ) = 0, \\ 
 \end{array}
\end{equation}

Next, we find wave solution of the anisotropic acoustic equation (14) in the annular layer, $R_1  \le R \le R_2$,  with interface continuous conditions (39) and (40)

${ \bold Statement \ 3:}$ Suppose that the point delta source (2) is in outside of whole sphere, $r_s  > R_2 $, 
the pressure acoustic wave $P(\vec R)$  is solution of the pressure acoustic equation (14) in the sphere annular layer domain $R_1  \le R \le R_2$ with anisotropic relative acoustic parameters in (15)-(17) or (20)-(22);, also, satisfies the interface continuous condition   (39) and (40) on the inner interface boundary $\vec R = \vec R_1 $, then
\begin{equation}
\begin{array}{l}
 P(\vec R) =  - \frac{1}{{4\pi }}\frac{{e^{ik_b \left| {\left( {Q^{ - 1} (R - R_1 ),\theta ,\phi } \right) - \mathord{\buildrel{\lower3pt\hbox{$\scriptscriptstyle\rightharpoonup$}} 
\over r} _s } \right|} }}{{\left| {\left( {Q^{ - 1} (R - R_1 ),\theta ,\phi } \right) - \vec r_s } \right|}} \\ 
  + \frac{1}{{4\pi }}ik_b j_0 \left( {k_b Q^{ - 1} (R - R_1 )} \right)h_0^{(1)} (k_b r_s ),   R_1  \le R \le R_2,\\
 \end{array}
\end{equation}
where $j_0$ is 0 order sphere Bessel function, $h_0^{(1)}$ zero order first Hankel function.  Moreover, on the outer interface boundary, acoustic wave $P(\vec R_2)$  is different from incident wave ${P_i}(\vec R_2)$ ,
\begin{equation}
\begin{array}{l}
 P(\vec R_2 ) =  - \frac{1}{{4\pi }}\frac{{e^{ik_b \left| {\left( {R_2 ,\theta ,\phi } \right) - \mathord{\buildrel{\lower3pt\hbox{$\scriptscriptstyle\rightharpoonup$}} 
\over r} _s } \right|} }}{{\left| {\left( {R_2 ,\theta ,\phi } \right) - \vec r_s } \right|}} \\ 
  + \frac{1}{{4\pi }}ik_b j_0 \left( {k_b R_2 } \right)h_0^{(1)} (k_b r_s ) \\ 
  = p_i (\vec R_2 ) + \frac{1}{{4\pi }}ik_b j_0 \left( {k_b R_2 } \right)h_0^{(1)} (k_b r_s ), \\ 
 \end{array}
\end{equation}
Proof: substitute anisotropic relative acoustic parameters in (15)-(17) into pressure acoustic equation (14),
\begin{equation}
\begin{array}{l}
 \frac{{\partial r}}{{\partial R}}\frac{\partial }{{\partial r}}\left( {Q^{ - 1} (R - R_1 )} \right)^2 \frac{{\partial R}}{{\partial r}}\frac{{\partial r}}{{\partial R}}\frac{{\partial P}}{{\partial r}} \\ 
  + \frac{{\partial r}}{{\partial R}}\frac{1}{{\sin \theta }}\frac{\partial }{{\partial \theta }}\sin \theta \frac{{\partial P}}{{\partial \theta }} \\ 
  + \frac{{\partial r}}{{\partial R}}\frac{1}{{\sin ^2 \theta }}\frac{{\partial ^2 P}}{{\partial \phi ^2 }} \\ 
  + \frac{{\partial r}}{{\partial R}}\left( {Q^{ - 1} (R - R_1 )} \right)^2 k_b ^2 P = 0, \\ 
 \end{array}
\end{equation}
Substitute inverse transformation (11) into (43), equation (43) is translated to
\begin{equation}
\begin{array}{l}
 \frac{\partial }{{\partial r}}r^2 \frac{{\partial P}}{{\partial r}} + \frac{1}{{\sin \theta }}\frac{\partial }{{\partial \theta }}\sin \theta \frac{{\partial P}}{{\partial \theta }} \\ 
  + \frac{1}{{\sin ^2 \theta }}\frac{{\partial ^2 P}}{{\partial \phi ^2 }} + r^2 k_b ^2 P = 0. \\ 
 \end{array}
\end{equation}

Substitute inverse transformation (11) into (41), the pressure acoustic wave $P(\vec R)$  to be
\begin{equation}
\begin{array}{l}
 P(\vec R) =  - \frac{1}{{4\pi }}\frac{{e^{ik_b \left| {\left( {Q^{ - 1} (R - R_1 ),\theta ,\phi } \right) - \mathord{\buildrel{\lower3pt\hbox{$\scriptscriptstyle\rightharpoonup$}} 
\over r} _s } \right|} }}{{\left| {\left( {Q^{ - 1} (R - R_1 ),\theta ,\phi } \right) - \vec r_s } \right|}} \\ 
 \left. { + \frac{1}{{4\pi }}ik_b j_0 \left( {k_b Q^{ - 1} (R - R_1 ),} \right)h_0^{(1)} (k_b r_s )} \right) \\ 
  =  - \frac{1}{{4\pi }}\frac{{e^{ik_b \left| {\vec R - \mathord{\buildrel{\lower3pt\hbox{$\scriptscriptstyle\rightharpoonup$}} 
\over r} _s } \right|} }}{{\left| {\vec R - \vec r_s } \right|}} \\ 
  + \frac{1}{{4\pi }}ik_b j_0 \left( {k_b R  } \right)h_0^{(1)} (k_b r_s ) \\ 
  = p_i (\vec R) + P_s (\vec R), \\ 
 \end{array}
\end{equation}
\begin{equation}
p_i (\vec R) =  - \frac{1}{{4\pi }}\frac{{e^{ik_b \left| {\vec R - \mathord{\buildrel{\lower3pt\hbox{$\scriptscriptstyle\rightharpoonup$}} 
\over r} _s } \right|} }}{{\left| {\vec R - \vec r_s } \right|}},
\end{equation}
\begin{equation}
P_s (\vec R) =  + \frac{1}{{4\pi }}ik_b j_0 \left( {k_b R } \right)h_0^{(1)} (k_b r_s ),
\end{equation}
                                 
Because the $p_i(\vec R)$  in (46) and $P_s(\vec R)$ in (47) in right hand side of (45) are solution of homogeneous acoustic wave equation (4) or (44) in the sphere $R\le R_2$  , therefore, pressure acoustic wave $P(\vec R)$ in (41), (i.e. in left hand side of (45)) does satisfy the acoustic equation (14) in the annular layer. By [5][7][8] and from (45) and (11)

\begin{equation}
\begin{array}{l}
 P(\vec R) = p_i (\vec R) + \frac{1}{{4\pi }}ik_b j_0 \left( {k_b Q^{ - 1} (R - R_1 )} \right)h_0^{(1)} (k_b r_s ) \\ 
  =  - \frac{1}{{4\pi }}\frac{{e^{ik_b \left| {\left( {Q^{ - 1} (R - R_1 ),\theta ,\phi } \right) - \mathord{\buildrel{\lower3pt\hbox{$\scriptscriptstyle\rightharpoonup$}} 
\over r} _s } \right|} }}{{\left| {\left( {Q^{ - 1} (R - R_1 ),\theta ,\phi } \right) - \vec r_s } \right|}} \\ 
  + \frac{1}{{4\pi }}ik_b j_0 \left( {k_b Q^{ - 1} (R - R_1 )} \right)h_0^{(1)} (k_b r_s ) \\ 
  = \sum\limits_{l = 1}^\infty  {} ik_b j_l (k_b Q^{ - 1} (R - R_1 ))h_l^{(1)} (k_b r_s ) \\ 
 \sum\limits_{m =  - l}^l {} Y_l^{(m)} (\theta ,\phi )Y_l^{(m)*} (\theta _s ,\phi _s ),Q^{ - 1} (R - R_1 ) \le r_s , \\ 
 \end{array}
\end{equation}

Because on $\vec R = \vec R_1 $.and $r_s > R_2 > R_1$ , when  $j \ge 1$,
\begin{equation}
\begin{array}{l}
 j_l (k_b Q^{ - 1} (R - R_1 ))|_{r = R_1 }  =  \\ 
  = j_l (k_b Q^{ - 1} (R_1  - R_1 )) = j_l (0) = 0,j \ge 1, \\ 
 \end{array}
\end{equation}
from (48)
\begin{equation}
\begin{array}{l}
 P(\vec R_1 ) =  \\ 
  =  - \sum\limits_{l = 1}^\infty  {} ik_b j_l (k_b Q^{ - 1} (R_1  - R_1 ))h_l^{(1)} (k_b r_s ) \\ 
 \sum\limits_{m =  - l}^l {} Y_l^{(m)} (\theta ,\phi )Y_l^{(m)*} (\theta _s ,\phi _s ) \\ 
  =  - \sum\limits_{l = 1}^\infty  {} ik_b j_l (0)h_l^{(1)} (k_b r_s ) \\ 
 \sum\limits_{m =  - l}^l {} Y_l^{(m)} (\theta ,\phi )Y_l^{(m)*} (\theta _s ,\phi _s ) = 0, \\ 
 \end{array}
\end{equation}             
the interface continuous condition (39),$P(\vec R_1^ +  ) = 0$, is satisfied. 

From equation (50) and transformation (7) and (11),
\begin{equation}
\begin{array}{l}
 \frac{\partial }{{\partial R}}P(\vec R) =  \\ 
  =  - \sum\limits_{l = 1}^\infty  {} ik_b \left( {\frac{l}{{(Q^{ - 1} (R - R_1 )}}\frac{{dr}}{{dR}}j_l (k_b Q^{ - 1} (R - R_1 ))} \right. \\ 
 \left. { - k_b \frac{{dr}}{{dR}}j_{l + 1} (k_b Q^{ - 1} (R - R_1 )} \right)h_l^{(1)} (k_b r_s ) \\ 
 \sum\limits_{m =  - l}^l {} Y_l^{(m)} (\theta ,\phi )Y_l^{(m)*} (\theta _s ,\phi _s ),Q^{ - 1} (r - R_1 ) \le r_s . \\ 
 \end{array}
\end{equation}                      
                      
From (40),
\begin{equation}   
\begin{array}{l}
 \left( {Q^{ - 1} (R - R_1 )} \right)^2 \frac{{dR}}{{dr}}\frac{\partial }{{\partial R}}P(\vec R)|_{\vec R = \vec R_1^ +  }  =  \\ 
  =  - \sum\limits_{l = 1}^\infty  {} ik_b \left( {lQ^{ - 1} (R - R_1 )j_l (k_b Q^{ - 1} (R - R_1 ))} \right. \\ 
 \left. { - k_b \left( {Q^{ - 1} (R - R_1 )} \right)^2 j_{l + 1} (k_b Q^{ - 1} (R - R_1 )} \right)h_l^{(1)} (k_b r_s ) \\ 
 \sum\limits_{m =  - l}^l {} Y_l^{(m)} (\theta ,\phi )Y_l^{(m)*} (\theta _s ,\phi _s )|_{R = R_1 } , \\ 
  =  - \sum\limits_{l = 1}^\infty  {} ik_b \left( {lQ^{ - 1} (R_1  - R_1 )j_l (k_b Q^{ - 1} (R_1  - R_1 ))} \right. \\ 
 \left. { - k_b \left( {Q^{ - 1} (R_1  - R_1 )} \right)^2 j_{l + 1} (k_b Q^{ - 1} (R_1  - R_1 )} \right)h_l^{(1)} (k_b r_s ) \\ 
 \sum\limits_{m =  - l}^l {} Y_l^{(m)} (\theta ,\phi )Y_l^{(m)*} (\theta _s ,\phi _s ) = 0, \\ 
 Q^{ - 1} (R - R_1 ) \le r_s , \\ 
 \end{array}
\end{equation}

therefore,
\begin{equation}   
\begin{array}{l}
 \frac{1}{{\rho _r }}R^2 \frac{\partial }{{\partial R}}P(\vec R_1 ^ +  ) \\ 
  = \left( {Q^{ - 1} (R - R_1 )} \right)^2 \frac{{dR}}{{dr}}\frac{\partial }{{\partial R}}P(\vec R)|_{\vec R = \vec R_1^ +  }  = 0, \\ 
 \end{array}
\end{equation}   

we have proved that the wave $P(\vec R)$ in (41) does satisfy the anisotropic acoustic equation (14) and interface zero wave value condition (39) and zero derivative condition (40) on the interface boundary $\vec R = \vec R_1$. The acoustic wave $P(\vec R)$ in (41) is 
caused by the condition that " the inner sphere is cloaked".
Moreover, on outer interface boundary  $\vec R = \vec R_2 $,  the value of the acoustic wave in (41) is,
\begin{equation}   
\begin{array}{l}
 P(\vec R_2 ) =  - \frac{1}{{4\pi }}\frac{{e^{ik_b \left| {\left( {Q^{ - 1} (R_2  - R_1 ),\theta ,\phi } \right) - \mathord{\buildrel{\lower3pt\hbox{$\scriptscriptstyle\rightharpoonup$}} 
\over r} _s } \right|} }}{{\left| {\left( {Q^{ - 1} (R_2  - R_1 ),\theta ,\phi } \right) - \vec r_s } \right|}} \\ 
 \left. { + \frac{1}{{4\pi }}ik_b j_0 \left( {k_b Q^{ - 1} (R_2  - R_1 ),} \right)h_0^{(1)} (k_b r_s )} \right) \\ 
  =  - \frac{1}{{4\pi }}\frac{{e^{ik_b \left| {\left( {R_2 ,\theta ,\phi } \right) - \mathord{\buildrel{\lower3pt\hbox{$\scriptscriptstyle\rightharpoonup$}} 
\over r} _s } \right|} }}{{\left| {\left( {R_2 ,\theta ,\phi } \right) - \vec r_s } \right|}} \\ 
  + \frac{1}{{4\pi }}ik_b j_0 \left( {k_b R_2 } \right)h_0^{(1)} (k_b r_s ) \\ 
  = p_i (\vec R_2) + P_s (\vec R_2), \\ 
 \end{array}
\end{equation}   
We have (42) 
\[
\begin{array}{l}
 P(\vec R_2 ) =  - \frac{1}{{4\pi }}\frac{{e^{ik_b \left| {\left( {R_2 ,\theta ,\phi } \right) - \mathord{\buildrel{\lower3pt\hbox{$\scriptscriptstyle\rightharpoonup$}} 
\over r} _s } \right|} }}{{\left| {\left( {R_2 ,\theta ,\phi } \right) - \vec r_s } \right|}} \\ 
  + \frac{1}{{4\pi }}ik_b j_0 \left( {k_b R_2 } \right)h_0^{(1)} (k_b r_s ) \\ 
  = p_i (\vec R_2 ) + \frac{1}{{4\pi }}ik_b j_0 \left( {k_b R_2 } \right)h_0^{(1)} (k_b r_s ),       (42) \\ 
 \end{array}
\]

Cloaked inner sphere causes the wave solution (41) of the anisotropic acoustic equation (14) is different from the incident wave on outer interface boundary $\vec R = \vec R_2 $. the necessary no scattering condition on the outer interface boundary $\vec R = \vec R_2$ is destroyed. Statement 3 is proved.

\section { $0$ to $R_1$ spherical radial transformation can not be used to induce acoustic no scattering cloak}

\subsection{  $0$ to $R_1$  spherical radial transformation can not be used to induce acoustic no scattering cloak}

${ \bold Statement \ 4:}$ For source incident acoustic wave or plane wave $p_i (\vec r)$ is in outside of whole sphere $r_s  \ge R_2 $, in annular layer domain, $R_1  \le R \le R_2$ , the anisotropic relative acoustic media  (15)-(17) or (20)-(22) is induced by  $0$ to $R_1$  spherical radial transformation (7) or (18), the background medium is in the inner sphere, if pressure acoustic wave $P_1(\vec R)$ is solution of acoustic equation (14) in annular layer domain,$R_1  \le R \le R_2$ , and on the interface  boundary  $\vec R = \vec R_2$, $P_1(\vec R)$  satisfies no scattering interface condition (26) and (27), then the inners sphere $R \le R_1$, can not be cloaked. Inversely, if the inners sphere $R \le R_1$,  is cloaked, then on the interface boundary $\vec R = \vec R_2$, the no scattering interface continuous condition (26) and (27) of the acoustic wave $P(\vec R)$, can not be satisfied, the whole sphere can cause scattering wave to disturb the incident wave and will be detected  Therefore, $0$ to $R_1$ spherical radial transformation can not be used to induce acoustic no scattering cloak. 

Proof:  For source incident acoustic wave $p_i (\vec r)$ is in outside of whole sphere, $r_s \ge R_2$, and the anisotropic acoustic media (15)-(17) in annular layer $R_1  \le R \le R_2$ is induced  by the $0$ to $R_1$  spherical radial transformation (7) or  (20)-(22)  by linear $0$ to $R_1$ spherical radial transformation (18), and background medium is in inner sphere. If  on the outer interface boundary $\vec R=\vec R_2$,  no scattering interface continuous conditions (26) and (27) is satisfied by  the wave solution  $P_1(\vec R)$   of the acoustic wave equation (14) in layer $R_1  \le R \le R_2$ , based on Statement 2, the acoustic wave $P_1(\vec R)$   is nonzero constant on the inner interface boundary  $\vec R=\vec R_1$, $P_1(\vec R_1 )  =  - \frac{1}{{4\pi }}\frac{{e^{ik_b \left| {r_s } \right|} }}{{\left| {r_s } \right|}} \ne 0$ in (37), then inner sphere $R \le R_1$,  can not be cloaked, otherwise, if inner sphere , $R \le R_1$, is cloaked , $P_2(\vec R)=0$, it does cause $P_1(\vec R_1)=0$,  that is contradiction with $P_1(\vec R_1 ) =  =  - \frac{1}{{4\pi }}\frac{{e^{ik_b \left| {r_s } \right|} }}{{\left| {r_s } \right|}} \ne 0$, that induces the acoustic wave equation in the inner sphere  with nonzero Dirichlet boundary value on the boundary $\vec R = \vec R_1$, solve the equation, the nonzero scattering acoustic wave propagation in the inner sphere $R \le R_1$, if
$j_0 (k_b R_1 ) \ne 0$,
\[
P_2 (\vec R) =  - \frac{1}{{4\pi }}\frac{{e^{ - ik_b r_s } }}{{r_s }}\frac{{j_0 (k_b R)}}{{j_0 (k_b R_1 )}}.
\]
The inner sphere $R \le R_1$ is not cloaked.
Inversely, if  inner sphere $R \le R_1$ is cloaked, $P_2(\vec R) =0$, in $R \le R_1$, it is obvious that wave  $P_2(\vec R) =0$,  satisfies background homogeneous acoustic equation (4) in inner sphere $R \le R_1$, moreover, the ¡°$P_2(\vec R) =0$, in $R \le R_1$, ¡±  that causes $P(\vec R_1)=0$ in (39) and its zero derivative condition in (40). Under the zero interface conditions (39) and (40), based on Statement 3,  on the outer interface boundary $\vec R=\vec R_2$ , the wave solution of the acoustic equation (14) is different from the incident wave $p_i(\vec R)$,
\[
\begin{array}{l}
 P(\vec R_2 ) =  - \frac{1}{{4\pi }}\frac{{e^{ik_b \left| {\left( {R_2 ,\theta ,\phi } \right) - \mathord{\buildrel{\lower3pt\hbox{$\scriptscriptstyle\rightharpoonup$}} 
\over r} _s } \right|} }}{{\left| {\left( {R_2 ,\theta ,\phi } \right) - \vec r_s } \right|}} \\ 
  + \frac{1}{{4\pi }}ik_b j_0 \left( {k_b R_2 } \right)h_0^{(1)} (k_b r_s )\;\;\;\;\;\;\;\;\;\;\;\;\;(42) \\ 
  = p_i (\vec R_2 ) + \frac{1}{{4\pi }}ik_b j_0 \left( {k_b R_2 } \right)h_0^{(1)} (k_b r_s ) \\
\ne {p_i (\vec R_2 ) },
 \\ 
 \end{array}
\]
that is basic contradiction with the necessary no scattering interface continuous condition on the outer interface boundary $\vec R=\vec R_2$ and (26)(27) , and for $l = 0$, the value of $P(\vec R_2)$ in (42) that induces the acoustic wave equation (1) (the zero order sphere Bessel equation) in outside of whole sphere,  $R \ge R_2$, with zero Dirichlet boundary value on the boundary $\vec R = \vec R_2$, solve the equation, the nonzero scattering acoustic wave propagation in the outside of the whole sphere, $R \ge R_2$, 
\[
p_s (\vec r) = \frac{{ik_b }}{{4\pi }}\frac{{j_0 (k_b R_2 )}}{{h_0^{(1)} (k_b R_2 )}}h_0^{(1)} (k_b r_s )h_0^{(1)} (k_b r),
\]
the incident wave in outside sphere can be disturbed, whole sphere can be detected, therefore, $0$ to $R_1$  spherical radial transformation can not be used to induce acoustic no scattering cloak. The statement 4 is proved.

\subsection{Pressure acoustic wave does satisfy the "no scattering interface continuous condition" (26) and (27) on the outer interface boundary $\vec R = \vec R_2$, the wave is propagation to penetrate into the inner sphere with background acoustic medium, if $j_1 (k_b R_1 ) = 0$    }

${ \bold Statement \ 5:}$ Suppose that the pressure acoustic wave $P_2(\vec R)$  is solution of the pressure acoustic equation (4) in the inner sphere $R \le R_1$,  with background acoustic speed medium  , moreover, the pressure acoustic wave $P_2(\vec R)$   does satisfy interface continuous conditions on the interface boundary $\vec R=\vec R_1$ 
\begin{equation}  
P_2(\vec R_1 ) =  - \frac{1}{{4\pi }}\frac{{e^{ik_b r_s } }}{{r_s }}, 
\end{equation}   
\begin{equation}  
\begin{array}{l}
 \frac{{\partial P_2(\vec R)}}{{\partial R}}|_{R = R_1 }  \\ 
  = \frac{{R_2 }}{{R_2  - R_1 }}(R - R_1 )^2 \frac{{\partial P_2}}{{\partial R}}|_{R = R_1 }  = 0, \\ 
 \end{array} 
\end{equation}                                    
If $j_1 (k_b R_1 ) = 0$, then bounded pressure acoustic wave
 \begin{equation} 
\begin{array}{l}
P_2(\vec R) = \frac{{k_b ^2 R_1 ^2 }}{{4\pi }}\frac{{e^{ik_b r{}_s} }}{{\left| {r_s } \right|}}n_1 (k_b R_1 )j_0 (k_b R) \\
=  - \frac{1}{{4\pi }}\frac{{e^{ - ik_b r_s } }}{{r_s }}\frac{{j_0 (k_b R)}}{{j_0 (k_b R_1 )}}.
 \end{array}  
\end{equation}
is propagation in inner sphere $R \le R_1$.                 
    
Proof: Because the interface continuous condition (55) and (56) are spherical symmetry,
the pressure wave $P_2(\vec R)$  in (57) is sphere symmetry function only depend on the radial
variable $R$,  $P_2(\vec R)$  in (57) does satisfy the following  0 order Sphere Bessel equation 
\begin{equation}
\frac{\partial }{{\partial R}}R^2 \frac{{\partial P_2}}{{\partial R}} + k_b ^2 R^2 P_2 = 0,   
\end{equation}  
\[ 
\begin{array}{l}
 P_2 (\vec R) = \frac{{k_b^2 R_1^2 }}{{4\pi }}\frac{{e^{ik_b r_s } }}{{r_s }}n_1 (k_b R_1 )j_0 (k_b R) \\ 
  =  - \frac{{k_b^2 R_1^2 }}{{4\pi }}\frac{{e^{ik_b r_s } }}{{r_s }}\left( {\frac{{\cos (k_b R_1 )}}{{(k_b R_1 )^2 }} + \frac{{\sin (k_b R_1 )}}{{k_b R_1 }}} \right)\frac{{\sin (k_b R)}}{{k_b R}} \\ 
  =  - \frac{{k_b^2 R_1^2 }}{{4\pi }}\frac{{e^{ik_b r_s } }}{{r_s }}\left( {\frac{{\cos (k_b R_1 )}}{{(k_b R_1 )^2 }}\frac{{\sin (k_b R)}}{{k_b R}} + \frac{{\sin (k_b R_1 )}}{{k_b R_1 }}\frac{{\sin (k_b R)}}{{k_b R}}} \right) \\ 
  =  - \frac{{k_b^2 R_1^2 }}{{4\pi }}\frac{{e^{ik_b r_s } }}{{r_s }}\frac{{\cos (k_b R_1 )}}{{(k_b R_1 )^2 }}\frac{{\sin (k_b R)}}{{k_b R}} \\ 
  - \frac{{k_b^2 R_1^2 }}{{4\pi }}\frac{{e^{ik_b r_s } }}{{r_s }}\frac{{\sin (k_b R_1 )}}{{k_b R_1 }}\frac{{\sin (k_b R)}}{{k_b R}}, \\ 
 \end{array}
\]
\[
\begin{array}{l}
 P_2 (\vec R) = \frac{{k_b^2 R_1^2 }}{{4\pi }}\frac{{e^{ik_b r_s } }}{{r_s }}n_1 (k_b R_1 )j_0 (k_b R)|_{R = R_1 }  \\ 
  =  - \frac{{k_b^2 R_1^2 }}{{4\pi }}\frac{{e^{ik_b r_s } }}{{r_s }}\left( {\frac{{\cos (k_b R_1 )}}{{(k_b R_1 )^2 }} + \frac{{\sin (k_b R_1 )}}{{k_b R_1 }}} \right)\frac{{\sin (k_b R_1 )}}{{k_b R_1 }} \\ 
  =  - \frac{{k_b^2 R_1^2 }}{{4\pi }}\frac{{e^{ik_b r_s } }}{{r_s }}\left( {\frac{{\cos (k_b R_1 )}}{{(k_b R_1 )^2 }}\frac{{\sin (k_b R_1 )}}{{k_b R_1 }} + \frac{{\sin (k_b R_1 )}}{{k_b R_1 }}\frac{{\sin (k_b R_1 )}}{{k_b R_1 }}} \right) \\ 
  =  - \frac{{k_b^2 R_1^2 }}{{4\pi }}\frac{{e^{ik_b r_s } }}{{r_s }}\frac{{\cos (k_b R_1 )}}{{k_b^2 R_1^2 }}\frac{{\sin ^2 (k_b R_1 )}}{{k_b R_1 }} \\ 
  - \frac{{k_b^2 R_1^2 }}{{4\pi }}\frac{{e^{ik_b r_s } }}{{r_s }}\frac{{\sin ^2 (k_b R_1 )}}{{k_b^2 R_1^2 }} \\ 
  =  - \frac{{k_b^2 R_1^2 }}{{4\pi }}\frac{{e^{ik_b r_s } }}{{r_s }}\frac{{\cos (k_b R_1 )}}{{k_b^2 R_1^2 }}\frac{{\sin ^2 (k_b R_1 )}}{{k_b R_1 }} \\ 
  - \frac{{k_b^2 R_1^2 }}{{4\pi }}\frac{{e^{ik_b r_s } }}{{r_s }}\left( {\frac{1}{{k_b^2 R_1^2 }} - \frac{{\cos ^2 (k_b R_1 )}}{{k_b^2 R_1^2 }}} \right) \\ 
  =  - \frac{1}{{4\pi }}\frac{{e^{ik_b r_s } }}{{r_s }} - \frac{{k_b^2 R_1^2 }}{{4\pi }}\frac{{e^{ik_b r_s } }}{{r_s }}\frac{{\cos (k_b R_1 )}}{{k_b R_1 }}j_1 (k_b R_1 ) \\ 
  =  - \frac{1}{{4\pi }}\frac{{e^{ik_b r_s } }}{{r_s }}, \\ 
 \end{array}
\]

Therefore, $P_2(\vec R)$  in (57) does satisfy acoustic equation (4) and the following interface continuous conditions on the interface boundary $\vec R=\vec R_1$ ,

\[
P_2(\vec R)|_{R = R_1 ^ -  }  =  - \frac{1}{{4\pi }}\frac{{e^{ik_b r{}_s} }}{{\left| {r_s } \right|}},        (55)
\]
                                           
and
\[
\frac{\partial }{{\partial R}}P_2(R)|_{r = R_1 }  = \frac{{k_b }}{{4\pi }}\frac{{e^{ik_b r{}_s} }}{{\left| {r_s } \right|}}j_1 (k_b R_1 ) = 0.     (56)
\]
                              
The Statement 5 is proved. 

Based on Statement 2 and Statement 5, If the pressure wave $P_1(\vec R)$  is the solution of the acoustic wave equation (14) and satisfy the necessary no scattering interface continuous conditions
(26) and (27) on the outer interface boundary $\vec R=\vec R_2$ , moreover, for infinity countable angular frequencies $\omega _m $ that make $j_1 (k_{b,m} R_1 ) = 0
$ , then  the $P_2(\vec R)$  is  propagation to penetrate into the inner sphere $R  \le  R_1$, 
\[
P_2(\vec R) =  - \frac{1}{{4\pi }}\frac{{e^{ik_b r{}_s} }}{{\left| {r_s } \right|}}j_0 (k_b R),      (57)
\]
The inner sphere can not be cloaked.

In the Statement 4, for $j_1 (k_b R_1 ) \ne 0$, we prove that the isotropic background medium in the inner sphere $R \le R_1$ and induced anisotropic acoustic media (15)-(17) or (20)-(22) in the annular layer $R_1 \le R \le R_2$ is inconsistent that does cause the anisotropic 
acoustic equation (14), the interface continuous conditions on the outer interface boundary $\vec R=\vec R_2$ and on the inner interface boundary $\vec R =\vec R_1$ are contradiction equation system
and there exist  no physical wave solution to satisfy the above global equations system if $j_1 (k_b R_1 ) \ne 0$,   . In statement 5, we prove that if $j_1 (k_b R_1 ) = 0$, then there exist acoustic wave solution to satisfy the above global acoustic equation system. The pressure acoustic wave $P_1(\vec R)$ does satisfy the anisotropic acoustic equation (14) and does satisfy the no scattering interface continuous  conditions (26) and (27) on
the outer interface  boundary $\vec R =\vec R_2$, then on the inner interface boundary $\vec R = \vec R_1$
the acoustic wave $P_1(\vec R)$ is nonzero constant and is propagation to penetrate into
the inner sphere and does satisfy acoustic equation and interface continuous  conditions on the interface boundary $\vec R =\vec R_1$, the inner sphere can not be cloaked.

\subsection{Pressure acoustic wave does satisfy the "no scattering interface  continuous condition" (26) and (27) on the outer interface  boundary $\vec R = \vec R_2$, the wave is propagation to penetrate into the inner sphere with novel acoustic anisotropic media }

In this section, we propose a novel anisotropic media in the sphere $R \le R_1$ that is the induced anisotropic acoustic media  $ (15) \ to \ (17) $ or $ (20) \ to \ (22) $ by $0$ to $R_1$ sphere radial linear transformation (7) or(18),respectively, The anisotropic acoustic medium formulas in the sphere $R \le R_1$ is same as that in the annular layer $R_1 \le R \le R_2$.We prove if the anisotropic acoustic media are installed in  the sphere  $R \le R_1$, and the acoustic wave solution of the acoustic wave equation (14) and does satisfy no scattering interface  continuous condition (26) and (27) on the outer interface  boundary $\vec R= vec R_2$, then the acoustic wave is continuous propagation to penetrate into the inner sphere $R \le R_1$, 
the inner sphere $R \le R_1$, can not be cloaked. The anisotropic acoustic wave equation (14) with anisotropic media $(20) \ to \ (22)$ is,
\begin{equation}  
\begin{array}{l}
 \frac{\partial }{{\partial R}}\frac{{R_2 (R - R_1 )^2 }}{{R_2  - R_1 }}\frac{{\partial P(\vec R)}}{{\partial R}} \\ 
  + \frac{{R_2 }}{{R_2  - R_1 }}\frac{1}{{\sin \theta }}\frac{\partial }{{\partial \theta }}\sin \theta \frac{{\partial P(\vec R)}}{{\partial \theta }} \\ 
  + \frac{{R_2 }}{{R_2  - R_1 }}\frac{1}{{\sin ^2 \theta }}\frac{{\partial ^2 P(\vec R)}}{{\partial \phi ^2 }} \\ 
  + \left( {\frac{{R_2 }}{{R_2  - R_1 }}} \right)^3 (R - R_1 )^2 k_b ^2 P(\vec R) = 0, \\ 
 \end{array}
\end{equation}

${ \bold Statement \ 6:}$ Suppose that anisotropic acoustic media (20)-(22) is installed in sphere annular layer $R_1  \le R \le R_2$ which is induced by linear transformation (18), the  anisotropic acoustic media by same formulas (20)-(22) is installed  in the sphere $R \le R_0$. Then the acoustic wave $P_1(\vec R)$ is solution of the acoustic equation (59) or (14) in the annular layer $R_1  \le R \le R_2$,  the wave $P_1(\vec R)$ and incident wave $p_i(\vec r)$ satisfy "no scattering interface  continuous conditions" (26) and (27) on the outer interface  boundary  $\vec R =\vec R_2$,  the acoustic wave $P_2(\vec R)$ is solution of the acoustic equation (59) or (14) in the inner sphere $R \le R_1$, the wave  $P_2(\vec R)$ and   $P_1(\vec R)$   satisfy the interface continuous conditions on the inner interface  boundary $\vec R = \vec R_1$, the acoustic wave $P_1(\vec R)$ by (29) in $R_1  \le R \le R_2$ annular layer is proved in statement 1,
\[  
\begin{array}{l}
 P_1(\vec R) = p(r(R),\theta ,\phi ) \\ 
  =  - \frac{1}{{4\pi }}\frac{{e^{ik_b \left| {(Q^{ - 1} (R - R_1 ),\theta ,\phi ) - \vec r_s } \right|} }}{{\left| {(Q^{ - 1} (R - R_1 ),\theta ,\phi ) - \vec r_s } \right|}} \\ 
  =  - \frac{1}{{4\pi }}\frac{{e^{ik_b \left| {(\frac{{R_2 }}{{R_2  - R_1 }}(R - R_1 ),\theta ,\phi ) - \vec r_s } \right|} }}{{\left| {(\frac{{R_2 }}{{R_2  - R_1 }}(R - R_1 ),\theta ,\phi ) - \vec r_s } \right|}},      (29) \\ 
 \end{array} 
\]
in inner sphere $R \le R_1$, acoustic wave $P_2(\vec R)$ is 
\begin{equation}  
P_2(\vec R) =  - \frac{1}{{4\pi }}\frac{{e^{ik_b r_s } }}{{r_s }}j_0 \left( {k_b \frac{{R_2 }}{{R_2  - R_1 }}(R_1  - R)} \right), 
\end{equation}                                    
by Statement 2, on the interface boundary $\vec R=\vec R_1$ , $P_1(\vec R)$  is nonzero constant in (35), the acoustic wave $P_1(\vec R)$ in (29)  and wave $P_2(\vec R)$ in (60) satisfy the following interface continuous conditions on the interface boundary $\vec R = \vec R_1$ :
\begin{equation}
P_2(\vec R_1^ ) =P_1(\vec R_1^ ) =  - \frac{1}{{4\pi }}\frac{{e^{ik_b \left| {r_s } \right|} }}{{\left| {r_s } \right|}} \ne 0,
\end{equation}  

\begin{equation}
\begin{array}{l}
\frac{{R_2 }}{{R_2  - R_1 }}(R - R_1 )^2 \frac{{\partial P_1}}{{\partial R}}|_{R = R_1 } \\
=\frac{{R_2 }}{{R_2  - R_1 }}(R - R_1 )^2 \frac{{\partial P_2}}{{\partial R}}|_{R = R_1 } = 0,   \\
\end{array} 
\end{equation}  
 the  acoustic wave $P_1(\vec R)$  and wave $P_2(\vec R)$   are propagation to penetrate into the sphere $R \le R_0$, therefore the inner sphere  $R \le R_0$ can not be cloaked, 

Proof: If the pressure acoustic wave $P_1(\vec R)$ in (29) is solution of the pressure acoustic equation (59) or (14) in the sphere layer $R_1  \le R \le R_2$  with induced anisotropic media (20)-(22) by linear $0$ to $R_1$ sphere radial transformation (18), and  wave $P_1(\vec R)$ and incident wave $p_i(\vec r)$  satisfy the necessary "no scattering interface  continuous conditions" (26) and (27) on the outer interface  boundary $\vec R = \vec R_2$, by Statement 1, then wave
 $P_1(\vec R)$ is formula by (29), by Statement 2, on the interface boundary $\vec R=\vec R_1$ , $P_1(\vec R)$  is nonzero constant in (35), 
\[
P_1(\vec R_1 ) =  - \frac{1}{{4\pi }}\frac{{e^{ik_b \left| {r_s } \right|} }}{{\left| {r_s } \right|}} \ne 0,\;\;\;\;\;\;\;(35)
\]  
and its radial derivative does satisfy
\begin{equation}
\frac{{R_2 }}{{R_2  - R_1 }}(R - R_1 )^2 \frac{{\partial P_1}}{{\partial R}}|_{R = R_1 }  = 0,   
\end{equation}

The first step of the proof is finished. In the Second step, because the interface continuous conditions (61) and (62) on the interface boundary $\vec R = \vec R_1$, the wave $P_2(\vec R)$ and its radial derivative are only constant and it is not depend on angular variable, therefore the anisotropic acoustic equation (59) or (14) in inner sphere $ R \le R_1$  becomes the one dimensional sphere Bessel equation
\begin{equation}
\begin{array}{l} 
\frac{\partial }{{\partial R}}\frac{{R_2 }}{{R_2  - R_1 }}(R - R_1 )^2 \frac{{\partial P_2}}{{\partial R}} \\
+ k_b ^2 \left( {\frac{{R_2 }}{{R_2  - R_1 }}} \right)^3 (R - R_1 )^2 P_2 = 0, \\  
\end{array}
\end{equation}  
                                                          
Substitute $P_2(\vec R) =  - \frac{1}{{4\pi }}\frac{{e^{ik_b r_s } }}{{r_s }}j_0 \left( {k_b \frac{{R_2 }}{{R_2  - R_1 }}(R - R_1 )} \right)$ in (60) into the above one dimensional sphere Bessel equation (64),
Let $r = \frac{{R_2 }}{{R_2  - R_1 }}(R - R_1 )$, the equation (64) becomes
\begin{equation}  
\begin{array}{l} 
\frac{{\partial r}}{{\partial R}}\frac{\partial }{{\partial r}}\frac{{R_2 }}{{R_2  - R_1 }}(R - R_1 )^2 \frac{{\partial r}}{{\partial R}}\frac{{\partial P_2}}{{\partial r}}\\
 + k_b ^2 \left( {\frac{{R_2 }}{{R_2  - R_1 }}} \right)^3 (R - R_1 )^2 P_2 = 0,
\\
\end{array}
\end{equation}  
\begin{equation} 
\begin{array}{l}  
\frac{{R_2 }}{{R_2  - R_1 }}\frac{\partial }{{\partial r}}\left( {\frac{{R_2 }}{{R_2  - R_1 }}(R - R_1 )} \right)^2 \frac{{\partial P_2}}{{\partial r}} \\
+ k_b ^2 \left( {\frac{{R_2 }}{{R_2  - R_1 }}} \right)^3 (R - R_1 )^2 P_2 = 0,
\ \
\end{array}
\end{equation}  
\[
\begin{array}{l}  
\frac{\partial }{{\partial r}}\left( {\frac{{R_2 }}{{R_2  - R_1 }}(R - R_1 )} \right)^2 \frac{{\partial P_2}}{{\partial r}} \\
 + k_b ^2 \left( {\frac{{R_2 }}{{R_2  - R_1 }}} \right)^2 (R - R_1 )^2 P_2 = 0,
 \ \
\end{array}
\]

\begin{equation}
\frac{\partial }{{\partial r}}r^2 \frac{{\partial P_2}}{{\partial r}} 
+ k_b ^2 r^2 P_2 = 0,  
\end{equation}  

Let $r = \frac{{R_2 }}{{R_2  - R_1 }}(R - R_1 )$, it is obvious that $P_2(\vec R)$ in (60) can be translated to
\begin{equation}   
\begin{array}{l}
 P_2(\vec R) =  - \frac{1}{{4\pi }}\frac{{e^{ik_b r_s } }}{{r_s }}j_0 \left( {k_b \frac{{R_2 }}{{R_2  - R_1 }}(R - R_1 )} \right) \\ 
  =  - \frac{1}{{4\pi }}\frac{{e^{ik_b r_s } }}{{r_s }}j_0 (k_b r), \\ 
 \end{array}
\end{equation}  
so, the $P_2(\vec R)$ in (60) or (69) is solution of the equation (68), therefore, $P_2(\vec R) =  - \frac{1}{{4\pi }}\frac{{e^{ik_b r_s } }}{{r_s }}j_0 \left( {k_b \frac{{R_2 }}{{R_2  - R_1 }}(R - R_1 )} \right)$ in (60) is solution of the acoustic equation (64),(59) or (14). Also
\begin{equation}
\begin{array}{l}
 P_2(\vec R)|_{\vec R = \vec R_1 }  \\ 
  =  - \frac{1}{{4\pi }}\frac{{e^{ik_b r_s } }}{{r_s }}j_0 \left( {k_b \frac{{R_2 }}{{R_2  - R_1 }}(R_1  - R_1 )} \right) \\ 
  =  - \frac{1}{{4\pi }}\frac{{e^{ik_b r_s } }}{{r_s }} \\
=P_1(\vec R)|_{\vec R = \vec R_1 }, \\   
 \end{array}   
\end{equation}  
\begin{equation}
\begin{array}{l} 
\frac{{R_2 }}{{R_2  - R_1 }}(R - R_1 )^2 \frac{{\partial P_2}}{{\partial R}}|_{R = R_1 }  = \\
=\frac{{R_2 }}{{R_2  - R_1 }}(R - R_1 )^2 \frac{{\partial P_1}}{{\partial R}}|_{R = R_1 }  = 0, \\
 \end{array}   
\end{equation}                                                     
Therefore pressure wave solution of (14) or (59),$P_1(\vec R)$ and $P_2(\vec R)$ satisfy the interface continuous conditions (60) and (61) on the interface boundary $\vec R=\vec R_1$ , the Second step of the proof is finished. 
Summary, in the statement 6, we proved that pressure wave solution of (14) or (59), $P_1(\vec R)$ and incident wave $p_i(\vec r)$ satisfy "no scattering
interface continuous conditions" (26) and (27) on outer interface  boundary $\vec R = \vec R_2$, wave solution of (14) or (59),$P_1(\vec R)$ and $P_2(\vec R)$ satisfy the interface continuous conditions (61) and (62) on the inner interface boundary $\vec R = \vec R_1$, 
the acoustic wave $P_1(\vec R)$ and $P_2(\vec R)$ are continuous propagation to penetrate into the inner sphere $R \le R_1$.
Therefore the inner sphere $R \le R_1$  can not be cloaked. The Statement 6 is proved. 

\section {Discussions}
Using $0$ to $R_1$  sphere radial linear transformation, Pendry et al proposed EM invisible cloak [3]. It is proved that by [5][6] Pendry EM cloak [3] is invisible cloak with infinite speed and exceeding light speed fundamental difficulties. However, in the paper [4], authors did use 0 to $R_1$  spherical radial linear  coordinate transformation to induce acoustic cloak, but we proved that is not "no scattering acoustic cloak". In the page 024103-4 of paper [4],  from interface boundary derivative continuous condition (28) in [4],  $C_n =0 $, for all $n = 0,1,2 \cdots $ was obtained. However, for $n=0$ , left hand side of (27) in [4], $K_0 j_0 [k_{sh} (r - a)]|_{r = a}  = K_0  \ne 0$ , the right hand of (27),$C_0 j_0 (k_0 r)|_{r = a}  = 0$  , So the basic interface  continuous condition equation (27) in [4] is not satisfied. Authors of [4] used interface boundary derivative continuous conditions (28)  to derive zero scattering in the inner sphere $R \le R_1$ (in $r \le a$ in [4]),which inversely destroyed wave field interface  continuous conditions (27) that caused contradiction equation (27). Authors in paper [4]  
 used  [ Physically, this is because the radial mass density tends to infinity at the inner edge of the shell which reduces all radial particle motion to zero ]  to explain equantion (28) in [4], but that  can not explain the basic contradiction equation (27) in [4]. The contradiction equation  (27)  in the paper [4] is caused by inconsistent between the relative induced anisotropic acoustic media (20)-(22) in layer $R_1  \le R \le R_2$  by 0 to $R_1$ (or  0 to a in [4]) linear transformation (18) and background medium in inner sphere $R \le R_1$  and  $j_1 (k_b R_1 ) \ne 0$ , (or $j_1 (k_0 a) \ne 0$ .in [4]). 
The inconsistent causes there exist no solution of the anisotropic acoustic equations and interface  continuous equations system.
In statement 5,  we proved to chose infinity countable suitable angular frequency $\omega $, inner sphere radius  $R_1$, and background acoustic speed $c_b$  , that  making $j_1 (k_b R_1 ) = j_1 (\frac{\omega }{{c_b }}R_1 ) = 0$ ,and the acoustic wave solution of the acoustic wave equation (14) and does satisfy "no scattering interface  continuous condition" (26) and (27) on the outer interface boundary $\vec R= vec R_2$,  then the  acoustic wave  $P_1(\vec R)$ is propagation to penetrate into the  inner sphere $R \le R_1$  and becomes to $P_2 (\vec R) =  - \frac{1}{{4\pi }}\frac{{e^{ik_b r{}_s} }}{{\left| {r_s } \right|}}j_0 (k_b r)$ in (57), and the $P_1(\vec R)$ and $P_2(\vec R)$ satisfy interface continuous conditions without contradiction equation in [4] , the inner sphere can not be cloaked.
note that the anisotropic media by $(20)\  to \ (22)$ in the annular layer
$R_1 \le R \le R_2$ that is same as anisotropic media in [4], we only add the condition  $j_1 (k_b R_1 ) = j_1 (\frac{\omega }{{c_b }}R_1 ) = 0$,and the acoustic wave solution of the acoustic wave equation (14) and does satisfy "no scattering interface  continuous condition" (26) and (27) on the outer interface boundary $\vec R= vec R_2$,  the acoustic wave is propagation and continuous to penetrate into 
the sphere $R \le R_1$ and without the contradiction equation [for example, (27) in [4]], Therefore,  explain of authors in paper [4] [ Physically, this is because the radial mass density tends to infinity at the inner edge of the shell which reduces all radial particle motion to zero ] that can not be used to explain the basic contradiction equation (27) in [4].  The acoustic cloak in [4] is not acoustic no scattering cloak, the inner sphere can not be cloaked. 
$0$ to $R_1$  sphere radial linear transformation can not be used to induce
acoustic no scattering cloak.
By $0$ to $R_1$  sphere radial transformation (7)-(10), the background sphere $r \le R_2$ is compressed into the sphere annular layer domain, $R_1  \le R \le R_2$ .A new sphere
$r \le R_1$ is expanded by inflated of ¡°zero¡±. Originally, A new sphere $r \le R_1$is absolute
empty space without physical acoustic media. We should install acoustic medium in the
new inner sphere $r \le R_1$. We proved that the induced anisotropic media $ (15) \ to \ (17)$ or $(20) \ to \ (22)$ in the annular layer $R_1 \le r \le R_2$ is inconsistent with the background acoustic speed medium  in the inner sphere $r \le R_1$. The inconsistent shows that the induced anisotropic media $ (15) \ to \ (17)$ or $(20) \ to \ (22)$ in the annular layer $R_1 \le r \le R_2$ is not "No scattering acoustic cloak" media. we prove that the acoustic cloak[4] is not "no scattering acoustic cloak", and we prove that the $0$ to $R_1$ spherical radial transformation method can not be used to induce acoustic no scattering cloak. 

In statement 6,we propose a novel anisotropic media in the sphere $R \le R_1$ that is the induced anisotropic acoustic media  $ (15) \ to \ (17) $ or $ (20) \ to \ (22) $ by $0$ to $R_1$ sphere radial linear transformation (7) or(18),respectively, The anisotropic acoustic medium formulas in the sphere $R \le R_1$ is same as that in the annular layer $R_1 \le R \le R_2$.We prove if the anisotropic acoustic media are installed in  the sphere  $R \le R_1$, and the acoustic wave solution of the acoustic wave equation (14) and does satisfy "no scattering interface continuous condition" (26) and (27) on the outer interface  boundary $\vec R= vec R_2$, then the acoustic wave solution of the acoustic wave equation (14) is continuous propagation to penetrate into the inner sphere $R \le R_1$, 
the inner sphere $R \le R_1$, can not be cloaked. In our statement 6, we install the anisotropic acoustic media $(20) \ - \ (22)$ in the annular 
layer $R_1 \le R \le R_2$ and inner sphere $R \le R_1$ that are same as $[(6)  - \ (8)$ in page 024301-2 of paper [4], the induced density
are going to infinity on the both sides of the interface boundary edge $\vec R = \vec R_1$, if by explain of authors in paper [4], the pressure $P(\vec R)$ should be zero in the inner sphere $R \le R_1$,  however, in statement 6, we  proved that the nonzero and bounded acoustic wave is continuous
propagation  to penetrate into the inner sphere $R \le R_1$, and in inner sphere $R \le R_1$, the nonzero acoustic wave $P_2(\vec R)$ is bounded and continuous solution of the acounstic equation (59) or (14) with the induced anisotropic acoustic media $(20) \ - \ (22)$,
\[
P_2(\vec R) =  - \frac{1}{{4\pi }}\frac{{e^{ik_b r_s } }}{{r_s }}j_0 \left( {k_b \frac{{R_2 }}{{R_2  - R_1 }}(R_1  - R)} \right),     (60)
\]        
therefore, the explain of the authors in [4] [ Physically, this is because the radial mass density tends to infinity at the inner edge of the shell which reduces all radial particle motion to zero ] that can not be used to explain the basic contradiction equation (27) in paper [4]. The physical explain in paper [4] is not physical base of the "no scattering acoustic cloak" in [4]. The cloak in paper [4]  is not "acoustic No scattering cloak", the inner sphere $R \le R_1$ [$( r \le a )$ in [4]] can not be cloaked.
 
In ststement 4, we proved that if the inner sphere is cloaked that causes the
interface continuous condition (26) and (27) can not be satisfied that is contradiction with the necessary "No Scattering interface continuous condition" (26) and (27) on the outer interface   boundary $\vec R = \Vec R_2$. Because  radial mass density does no tend to infinity at the outer edge of the shell, $\vec R = \Vec R_2$', the physical
explain in [4] is fail to explain the contradiction in outer interface  boundary,  $\vec R = \Vec R_2$'. These undisputed evidences and theoretical proofs  prove that the physical explain in paper [4] is not physical base of the "no scattering acoustic cloak" in [4]. The cloak in paper [4]  is not "acoustic No scattering cloak", the inner sphere $R \le R_1$ [$( r \le a )$ in [4]] can not be cloaked.

0 to $R_1$ spherical radial transformation can not be used to induce acoustic no scattering cloak. We have proposed the Dirichelet to Neumann boundary value nonliear operator method and one dimensional well posed acoustic wave coefficient inversion in 1986 [10] and three dimensional acoustic wave scattering inversion in 1988 [11]. Based on our
acoustic wave velocity inversion, without transform, we proposed  GILD and GL no scattering method [12],[13] and propose GLLH and GLHUA double layer electromagnetic invisible cloak without exceeding light speed violation propagation [1][2][5][7], which mehod can be used to make acoustic no scattering cloak in next paper will be submitted.

Finaly, we discuss the case $k_b = 0 $ that can be caused by the zero frequency or infinity background acoustic speed $c_b$. When $k_b = 0 $, the acoustic 
equation becomes to the variable coefficient homogeneous Laplace-Betrami equation
In [14], author consider static electric conductivity equation in absence source, that is variable coefficient homogeneous Laplace-Betrami equation. 1 to 0 (D-to-N) is boundary map of any variable coefficient Laplace-Betrami equation in absence source. The no uniqueness of this inverse scattering problem is obvious, Counterexamples in [14] is no necessary and trivial for variable coefficient Laplace-Betrami equation. Transformation hole in [14] is not "acoustic no scattering cloak", because for any space domain with background material or any continuation material hole, unity field 1 is always obvious solution of variable coefficient Laplace-Betrami  equation in absence source. The field 1 is always to penetrate into the hole if boundary D to N map is 1 to 0. Inversely, to cloak hole that causes the boundary map to become 0 to 0, causes scattering field to disturb incident map 1 to 0 to map 0 to 0, the whole domain is detected and exposed,  Counterexamples in paper [14] is not ¡°acoustic no scattering cloak", Counterexamples in paper [14] is not relative to the elctromagnetic invisible cloak in [1][2][3][5][7].

\section {Physical Letter }

We propose our Global and Local field method and novel approach to prove the cloak in paper [4] is not "No scattering acoustic cloak". First, we define "the global acoustic equation system" , Second, we difine "acoustic no scattering cloak" as follows that suppose that the anisotropic acoustic media is created in the annular layer that make there exist acoustic wave solution to satisfy the global acoustic equation system, if in the outside of whole sphere the acoustic wave solution equal to incident wave, $p(\vec r) = p_i(\vec r)$, i.e. there exist no scattering wave to disturb the incident wave, and the acoustic wave solution is zero in the inner sphere, i.e. the inner sphere is cloaked, then the annular layer,  $R_1 \le R \le R_2$   and inner sphere $R \le R_1$  is called the "acoustic no scattering cloak. In Statement 7 in this paper, we prove if the induced anisotropic acoustic media (73) - (75) (i.e. (25),(24), (26) in paper [4]) by $0$ to $R_1$ linear transformation "0R1SRLT" in (71)-(22) is installed in the annular layer
$R_1 \le R \le R_2$, and $j_1 (k_b R_1 ) \ne 0$, then there exist no acoustic wave solution
to satisfy the "global acoustic wave eqution system" in (61) to (68). That prove that the cloak in paper [4] is not "acoustic No scattering Cloak".
In the statement 8 in this paper, we prove if the induced anisotropic acoustic media (73) - (75) (i.e. (25),(24),(26) in paper [4]) by $0$ to $R_1$ linear transformation "0R1SRLT" in (71)-(72) is installed in the annular layer
$R_1 \le R \le R_2$, and $j_1 (k_b R_1 ) = 0$, then there exist acoustic wave solution
to satisfy the "global acoustic wave eqution system" in (61) to (68), the nonzero and continuous bounded acoustic wave solution is propagation to penetrate into the inner sphere, the
inner sphere $R \le R_1 $ is not cloaked. That prove that the cloak in paper [4] is not "acoustic No scattering Cloak". 
In the novel statement 9, we prove if the induced anisotropic acoustic media (73) - (75) by $0$ to $R_1$ linear transformation "0R1SRLT" in (71)-(72) is installed in the annular layer
$R_1 \le R \le R_2$ and the inner sphere $R \le R_1$, then there exist acoustic wave solution
to satisfy the "global acoustic wave eqution system" in (61) to (65) and (91) to (93), the nonzero and continuous bounded physical acoustic wave solution is propagation to penetrate into the inner sphere, $R \le R_1 $, the
inner sphere $R \le R_1 $ is not cloaked. That prove that the cloak in paper [4] is not "acoustic No scattering Cloak". 
 That prove that the "Physically ....." explantion in paper [4] is not correct and is a mistake in "acoustic No Scattering Cloak Super Physical Sciences". We proved that  0 to $R_1$ spherical radial transformation can not be used to induce acoustic no scattering cloak. The 0 to R1 spherical radial transformation can not be used to induce static electric conductivity no scattering cloak.
We define "the global acoustic equation system" as follows, chose $R_2 > R_1 > 0$, a whole sphere $R \le R_2$ is located in the 3D  full space,  the 3D full Space is split into the outside of the whole sphere $R \ge R_2$ with background isotropic acoustic media, the annular layer $R_1 \le R \le R_2$ with anisotropic acoustic media, the inner sphere $R \le R_1$ with background isotropic acoustic media,and two interfaces $\vec R = \vec R_2$ and $\vec R = \vec R_1$,  
The isotropic acoustic equation in the outside of the whole sphere, $R \ge R_2$
\begin{equation}
\begin{array}{l}
 \frac{\partial }{{\partial r}}\left( {r^2 \frac{{\partial p(\vec r)}}{{\partial r}}} \right) + \frac{1}{{\sin \theta }}\frac{\partial }{{\partial \theta }}\sin \theta \frac{{\partial p(\vec r)}}{{\partial \theta }} \\ 
  + \frac{1}{{\sin ^2 \theta }}\frac{{\partial ^2 p(\vec r)}}{{\partial \phi ^2 }} + k_b^2 r^2 p(\vec r) = S(\vec r,\vec r_s ), \\ 
 \end{array}
\end {equation}  
Sommerfeld radiation condition equation
\begin{equation}
\mathop {\lim }\limits_{r \to \infty } r(\partial p/\partial r - ikp(\vec r)) = 0,
\end{equation}            
where $p(\vec r)$ is the pressure wave field in the background space. $c_b$ is the background constant acoustic speed,$\omega $ is the angular frequency, let $k_b  = \frac{\omega }{{c_b }}$, $S(\vec r,\vec r_s )$ is the acoustic source located in $\vec r_s $, $r_s > R_2$.
The anisotropic acoustic equation is in the annular layer  $R_1 \le R \le R_2$,
\begin{equation}
\begin{array}{l}
 \frac{\partial }{{\partial R}}\frac{1}{{\rho _r }}R^2 \frac{{\partial P_1(\vec R)}}{{\partial R}} + \frac{1}{{\rho _\theta  \sin \theta }}\frac{\partial }{{\partial \theta }}\sin \theta \frac{{\partial P_1 (\vec R)}}{{\partial \theta }} \\ 
  + \frac{1}{{\rho _\phi  \sin ^2 \theta }}\frac{{\partial ^2 P_1 (\vec R)}}{{\partial \phi ^2 }} + \frac{1}{{\lambda }}k_b ^2 R^2 P_1(\vec R) = 0, \\ 
 \end{array}
\end{equation}   
Pressure interface continuous equation on $\vec R = \vec R_2$,
\begin {equation}
P_1(\vec R_2 ^ -  ) = p(\vec R_2 ^ +  ),
\end{equation}
velocity interface continuous equation on $\vec R = \vec R_2$,
\begin{equation}
\frac{1}{{\rho _r }} \frac{\partial }{{\partial R}}P_1(\vec R_2 ^ -  ) = \frac{\partial }{{\partial r}}p(\vec R_2 ^ +  ),
\end{equation}
The acoustic wave equation in the inner sphere $R \le R_1$
\begin{equation}
\begin{array}{l}
 \frac{\partial }{{\partial R}}\left( {R^2 \frac{{\partial P_2(\vec R)}}{{\partial R}}} \right) + \frac{1}{{\sin \theta }}\frac{\partial }{{\partial \theta }}\sin \theta \frac{{\partial P_2(\vec R)}}{{\partial \theta }} \\ 
  + \frac{1}{{\sin ^2 \theta }}\frac{{\partial ^2 P_2(\vec R)}}{{\partial \phi ^2 }} + k_b ^2 R^2 P_2(\vec R) = 0, \\ 
 \end{array}
\end{equation}
Pressure interface continuous equation on $\vec R = \vec R_1$,
\begin {equation}
P_2(\vec R_1 ^ -  ) = P_1(\vec R_1 ^ +  ),
\end{equation}
velocity interface continuous equation on $\vec R = \vec R_1$,
\begin{equation}
\frac{\partial }{{\partial R}}P_2(\vec R_1 ^ -  ) = \frac{1}{{\rho _r }} \frac{\partial }{{\partial r}}P_1(\vec R_1 ^ +  ).
\end{equation}
The equations (71) to (78) compose to the "global acoustic equation system". 
The source in (71) is
\begin {equation}
      S(\vec r,\vec r_s ) = \delta (r - r_s )\delta (\theta  - \theta _s )\delta (\phi  - \phi _s )/{\sin \theta },
\end {equation}
In the 3D background media full space,  the solution of the equation (71) with source (79) is called the incident acoustic wave $p_i (\vec r)$ ,
\begin {equation}
p_i (\vec r) =  - 0.25e^{ik_b \left| {\vec r - \vec r_s } \right|} /\left| {\vec r - \vec r_s } \right|/\pi. 
\end {equation}
We define "acoustic no scattering cloak" as follows:
Suppose that the anisotropic acoustic media is created in the annular layer that make there exist acoustic wave solution to satisfy the global acoustic equation system, if in the outside of whole sphere the acoustic wave solution equal to incident wave, $p(\vec r) = p_i(\vec r)$, i.e. there exist no scattering wave to disturb the incident wave, and the acoustic wave solution is zero in the inner sphere, i.e. the inner sphere is cloaked, then the annular layer,  $R_1 \le R \le R_2$   and inner sphere $R \le R_1$  is called the "acoustic no scattering cloak", the anisotropic acoustic media is called no scattering materials, the annular layer is called no scattering cloaking layer, the inner sphere is called no scattering cloaked concealment.  \hfill \break
$0$ to $R_1$ sphere radial coordinate linear transformation "0R1SRLT" is
\begin{equation}
R = R_1  + {(R_2  - R_1)r }/{R_2 },
\end{equation}
and its inverse  transformation
\begin{equation}
r  = {R_2 (R - R_1 )}/{(R_2  - R_1) },
\end{equation}
The induced anisotropic acoustic media by the above "0R1SRLT" in (81)-(82) is
\begin{equation}
\rho _r   = {(R_2  - R_1 )R^2 }/{(R_2(R - R_1 )^2)},
\end{equation}
\begin{equation}
\rho _\theta   = \rho _\phi   = {(R_2  - R_1)}/{R_2 },
\end{equation}
\begin{equation}
\lambda  = {{(R_2  - R_1 )^3}{R^2} }/{({{R_2}^3}(R - R_1 )^2)},
\end{equation}
${\bold Statement \ 1:}$  
If the induced anisotropic acoustic media (83) - (85) by $0$ to $R_1$ linear transformation "0R1SRLT" in (81)-(82) is installed in the annular layer
$R_1 \le R \le R_2$, and $j_1 (k_b R_1 ) \ne 0$, then there exist no acoustic wave solution
to satisfy the "global acoustic wave eqution system" in (71) to (78).
\\
${\bold Proof:}$ To solve (71) to (75) in the global acoustic equation system, the acoustic wave solution equal to incident wave, $p(\vec r)=p_i(\vec R)$ in outside of whole sphere, $R \ge R_2$,  by the statement 1 in our paper [15] in
arXiv:1706.05375v7, the  acoustic wave solution $P_1(\vec R)$ in the annular layer $R_1 \le R \le R_2$, is
\begin{equation}
\begin{array}{l}
 P_1(\vec R) = p_i(r(R),\theta ,\phi )  
  =  - \frac{1}{{4\pi }}\frac{{e^{ik_b \left| {(\frac{{R_2 }}{{R_2  - R_1 }}(R - R_1 ),\theta ,\phi ) - \vec r_s } \right|} }}{{\left| {(\frac{{R_2 }}{{R_2  - R_1 }}(R - R_1 ),\theta ,\phi ) - \vec r_s } \right|}}, \\ 
 \end{array}
\end{equation}
on the inner boundary $\vec R = \vec R_1$, the pressure acoustic wave $P_1(\vec {R_1}^+)$ is nonzero constant,
\begin{equation}
P_1(\vec {R_1}^+ ) =  - \frac{1}{{4\pi }}\frac{{e^{ik_b \left| {r_s } \right|} }}{{\left| {r_s } \right|}} \ne 0.
\end{equation}
and velocity is zero. Because on the inner bundary $\vec R = \vec R_1$,the pressure $P_1(\vec R_1 )$ is localized to nonzero costant in (87), and the velocity is
localized to zero, which are independent with the angular $\theta$ and $\phi$,
the pressure interface continuous equation on the interface $\vec R = \vec R_1$,(77) becomes
\begin {equation}
P_2( R_1 ^ -  ) = P_2(\vec R_1 ^ -  ) = P_1(\vec R_1 ^ +  )=  - \frac{1}{{4\pi }}\frac{{e^{ik_b \left| {r_s } \right|} }}{{\left| {r_s } \right|}} ,
\end{equation}
the fluid velocity interface continuous equation on the interface $\vec R = \vec R_1$,(78) becomes
\begin{equation}
\frac{\partial }{{\partial R}}P_2( R_1 ^-) =
\frac{\partial }{{\partial R}}P_2(\vec R_1 ^ -  ) = \frac{1}{{\rho _r }} \frac{\partial }{{\partial r}}P_1(\vec R_1 ^ +  )=0.
\end{equation}
The acoustic wave equation is the inner sphere $R \le R_1$ (76) becomes sphere
Bessel equation,
\begin{equation}
 \frac{\partial }{{\partial R}}\left( {R^2 \frac{{\partial P_2(R)}}{{\partial R}}} \right) + k_b ^2 R^2 P_2(R) = 0. \\ 
\end{equation}
The global acoustic equation system (71) to (78) is localized to local equation
system (88) to (90) in the inner sphere $R \le R_1$,the unique solution of equation (90) with Neumann boundary condition (89),
\begin{equation}
P_2 (\vec R) = 0,
\end{equation}
does not satisfy the pressure wave continuos equation (88), the inner sphere
can not be cloaked. Also, because $j_1 (k_b R_1 ) \ne 0$,  the unique solution of equation (90) with Dirichlet boundary condition (88), if ${j_0 (k_b R_1 )} \ne 0$,
\begin{equation}
P_2 (\vec R) =  - \frac{1}{{4\pi }}\frac{{e^{ - ik_b r_s } }}{{r_s }}\frac{{j_0 (k_b R)}}{{j_0 (k_b R_1 )}}.
\end{equation}
does not satisfy the velocity interface continuous equation (89),
 there exsit no
acoustic wave solution to satisfy the local acoustic equation system (88) to (90) in the inner sphere, therefore, there exist no acoustic wave solution to satisfy the global acoustic system (71) to (78), the inner sphere $R \le R_1$ can not be cloaked. 
Inversely, suppose that the inner sphere $R \le R_1$ is cloaked,
$P_2(\vec R) = 0 $, by the statement 3 in [15], the pressure acoustic wave $P_2(\vec R)=0$, in $R \le R_1$ and $P_1(\vec R)$ in the $R_1  \le R \le R_2$
\begin{equation}
\begin{array}{l}
 P_1(\vec R) =  - \frac{1}{{4\pi }}\frac{{e^{ik_b \left| {\left( {{(R_2/(R_2-R_1))} (R - R_1 ),\theta ,\phi } \right) - \mathord{\buildrel{\lower3pt\hbox{$\scriptscriptstyle\rightharpoonup$}} 
\over r} _s } \right|} }}{{\left| {\left( {{(R_2/(R_2-R_1))} ,\theta ,\phi } \right) - \vec r_s } \right|}} \\ 
  + \frac{1}{{4\pi }}ik_b j_0 \left( {k_b {((R_2/(R_2-R_1))}  (R - R_1 )} \right)h_0^{(1)} (k_b r_s ),   \\
 \end{array}
\end{equation}
satisfy the equations (76), (77) and (78) in the global acoustic equation system,
where $j_0$ is 0 order sphere Bessel function, $h_0^{(1)}$ zero order first Hankel function. Moreover, on the outer interface boundary, acoustic wave $P_1(\vec R_2)$  is different from incident wave ${p_i}(\vec R_2)$ ,
\begin{equation}
\begin{array}{l}
 P_1(\vec R_2 ) =  - \frac{1}{{4\pi }}\frac{{e^{ik_b \left| {\left( {R_2 ,\theta ,\phi } \right) - \mathord{\buildrel{\lower3pt\hbox{$\scriptscriptstyle\rightharpoonup$}} 
\over r} _s } \right|} }}{{\left| {\left( {R_2 ,\theta ,\phi } \right) - \vec r_s } \right|}} \\ 
  + \frac{1}{{4\pi }}ik_b j_0 \left( {k_b R_2 } \right)h_0^{(1)} (k_b r_s ) \\ 
  = p_i (\vec R_2 ) + \frac{1}{{4\pi }}ik_b j_0 \left( {k_b R_2 } \right)h_0^{(1)} (k_b r_s ), \\ 
 \end{array}
\end{equation}
Substitute $P_1(\vec R_2 ) $ in (94) into the interface continuous equation (74) and (75), the equation (74) is reduced to 
\begin {equation}
p_s(R_2)= \frac{1}{{4\pi }}ik_b j_0 \left( {k_b R_2 } \right)h_0^{(1)} (k_b r_s ), 
\end{equation}
and the equation (75) is reduced to 
\begin{equation}
\frac{\partial }{{\partial r}}p_s(\vec R_2 ^ +  )=
-\frac{1}{{4\pi }}i(k_b)^2 j_1 \left( {k_b R_2 } \right)h_0^{(1)} (k_b r_s ), 
\end{equation}
The acoustic wave equation (71) in the outside of whole sphere $R \ge R_2$ is reduced to the scattering sphere Bessel equation,
\begin{equation}
 \frac{\partial }{{\partial r}}\left( {r^2 \frac{{\partial p_s(r)}}{{\partial r}}} \right) + k_b ^2 r^2 p_s(r) = 0. \\ 
\end{equation}
the Sommerfeld radiation equation (72) is reduced to
\begin{equation}
\mathop {\lim }\limits_{r \to \infty } r(\partial p_s/\partial r - ikp_s( r)) = 0,
\end{equation}   
The global acoustic equation system (71)  to (78) is localized to local scattering equation
system (95) to (98) in the outside of whole sphere $R \ge R_2$, the unique solution of scattering equation (97) and (98) with Neumann boundary condition (96),
\begin{equation}
p_s (r) = \frac{{ik_b }}{{4\pi }}\frac{{h_0^{(1)} (k_b r)}}{{h_1^{(1)} (k_b R_2 )}}j_1 (k_b R_2 )h_0^{(1)} (k_b r_s ),
\end{equation}
does not satisfy the equation (95), the unique solution of scattering equation (97) and (98) with Dirichlet boundary condition (95)
\begin{equation}
p_s (r) = \frac{{ik_b }}{{4\pi }}\frac{{h_0^{(1)} (k_b r)}}{{h_0^{(1)} (k_b R_2 )}}j_0 (k_b R_2 )h_0^{(1)} (k_b r_s ),
\end{equation}
does not satisfy the equation (96).
There exist no scattering wave solution to satisfy the scattering equation system (95) to (98), therefore, there exist no acoustic wave solution to satisfy the global acoustic system equations ( (71)  to (78)), the statement 1 is proved. The statement 1 shows that the induced anisotropic acoustic media (83)
to (85) by linear transformation "0R1SRLT" (81) and (82) in annular layer is inconsistent with background acoustic media in inner sphere. the $0$ to $R_1$ sphere radial linear transformation can not be used to induce the acoustic no scattering cloak.
\hfill\break
${\bold Statement \ 2:}$  
If the induced anisotropic acoustic media (83) - (85) by $0$ to $R_1$ linear transformation "0R1SRLT" in (81)-(82) is installed in the annular layer
$R_1 \le R \le R_2$, and $j_1 (k_b R_1 ) = 0$, then there exist acoustic wave solution
to satisfy the "global acoustic wave eqution system" in (71) to (78), the nonzero and continuous bounded acoustic wave solution is propagation to penetrate into the inner sphere, the
inner sphere $R \le R_1 $ is not cloaked.
\\
${\bold Proof:}$ To repeat the proof in first paragraph of the statement 1,  Because on the inner bundary $\vec R = \vec R_1$, the pressure $P_1(\vec R_1 )$ is nonzero costant in (87), and velocity is zero, which are independent with the angular $\theta$ and $\phi$, the global acoustic equation system (71) to (78) is localized to local equation system (88) to (90) in the inner sphere $R \le R_1$, because $j_1 (k_b R_1 ) = 0$,  the unique solution of equation (90) with Dirichlet boundary condition (88), if ${j_0 (k_b R_1 )} \ne 0$,
\[
P_2 (\vec R) =  - \frac{1}{{4\pi }}\frac{{e^{ - ik_b r_s } }}{{r_s }}\frac{{j_0 (k_b R)}}{{j_0 (k_b R_1 )}}.     \ \ \ \ \ (92)
\]
does satisfy the velocity interface continuous equation (89),
 there exsit 
acoustic wave solution to satisfy the local acoustic equation system (88) to (90) in the inner sphere, therefore, there exist  acoustic wave solution to satisfy the global acoustic system (71) to (78), the nonzero acoustic wave $P_2 (\vec R)$ is propagation in the inner sphere, the inner sphere $R \le R_1$ is not cloaked. \hfill \break
The following statement is novel that the induced anisotropic acoustic media (83) - (85) by $0$ to $R_1$ linear transformation "0R1SRLT" in (81)-(82) is installed in the annular layer
$R_1 \le R \le R_2$ and the inner sphere $R \le R_1$, the equation (76) should be changed to the anisotropic acoustic equation in the annular layer  $ R \le R_1$,
\begin{equation}
\begin{array}{l}
 \frac{\partial }{{\partial R}}\frac{1}{{\rho _r }}R^2 \frac{{\partial P_2(\vec R)}}{{\partial R}} + \frac{1}{{\rho _\theta  \sin \theta }}\frac{\partial }{{\partial \theta }}\sin \theta \frac{{\partial P_2 (\vec R)}}{{\partial \theta }} \\ 
  + \frac{1}{{\rho _\phi  \sin ^2 \theta }}\frac{{\partial ^2 P_2 (\vec R)}}{{\partial \phi ^2 }} + \frac{1}{{\lambda }}k_b ^2 R^2 P_2(\vec R) = 0, \\ 
 \end{array}
\end{equation}   
the equation (77) is no changed,
the pressure interface continuous equation on the interface $\vec R = \vec R_1$,
\begin {equation}
P_2(\vec R_1 ^ -  ) = P_1(\vec R_1 ^ +  ),
\end{equation}
the equation (78) should be changed to 
fluid velocity interface continuous equation on the interface $\vec R = \vec R_1$,
\begin{equation}
\frac{1}{{\rho _r }(\vec R_1 ^ -  ) } \frac{\partial }{{\partial R}}P_2(\vec R_1 ^ -  ) = \frac{1}{{\rho _r }(\vec R_1 ^ +  ) } \frac{\partial }{{\partial r}}P_1(\vec R_1 ^ +  ).
\end{equation}
${ \bold Statement \ 3:}$ If the induced anisotropic acoustic media (83) - (85) by $0$ to $R_1$ linear transformation "0R1SRLT" in (81)-(82) is installed in the annular layer
$R_1 \le R \le R_2$ and the inner sphere $R \le R_1$, then there exist acoustic wave solution
to satisfy the "global acoustic wave eqution system" in (71) to (75) and (101) to (103), the nonzero and continuous bounded acoustic wave solution is propagation to penetrate into the inner sphere, $R \le R_1 $ 
\begin{equation}  
P_2( R) =  - \frac{1}{{4\pi }}\frac{{e^{ik_b r_s } }}{{r_s }}j_0 \left( {k_b \frac{{R_2 }}{{R_2  - R_1 }}(R- R_1 )} \right), 
\end{equation} 
the inner sphere $R \le R_1 $ is not cloaked. \hfill \break
 ${\bold Proof:}$ To repeat the proof in first paragraph of the statement 1, Because on the inner bundary $\vec R = \vec R_1$, the pressure $P_1(\vec R_1 )$ is nonzero costant in (87), and velocity is zero, which are independent with the angular $\theta$ and $\phi$, the "global acoustic wave eqution system"  (71) to (75) and (101) to (103),is localized to the following local equation system (105) to (107) in the inner sphere $R \le R_1$,  the equation (101) is reduced to the 1 - D sphere Bessel equation (105) with anistropic media,
\begin{equation}
\begin{array}{l} 
\frac{\partial }{{\partial R}}\frac{{R_2 }}{{R_2  - R_1 }}(R - R_1 )^2 \frac{{\partial P_2}}{{\partial R}} \\
+ k_b ^2 \left( {\frac{{R_2 }}{{R_2  - R_1 }}} \right)^3 (R - R_1 )^2 P_2 = 0, \\  
\end{array}
\end{equation} 
the equation (102) is reduced to the equation (106) 
\begin{equation}
P_2( R_1^ - )=P_2(\vec R_1^ - ) =P_1(\vec R_1^ + ) =  - \frac{1}{{4\pi }}\frac{{e^{ik_b \left| {r_s } \right|} }}{{\left| {r_s } \right|}},
\end{equation}  
the equation (103) is reduced to the equation (107)  
\begin{equation}
\begin{array}{l}
\frac{{R_2 }}{{R_2  - R_1 }}(R - R_1 )^2 \frac{{\partial P_2}}{{\partial R}}|_{R = R_1 }  \\
=\frac{{R_2 }}{{R_2  - R_1 }}(R - R_1 )^2 \frac{{\partial P_1}}{{\partial R}}|_{R = R_1 }  = 0,   \\
\end{array} 
\end{equation}
Let $r = \frac{{R_2 }}{{R_2  - R_1 }}(R - R_1 )$, the equation (105) becomes
\begin{equation}  
\begin{array}{l} 
\frac{{\partial r}}{{\partial R}}\frac{\partial }{{\partial r}}\frac{{R_2 }}{{R_2  - R_1 }}(R - R_1 )^2 \frac{{\partial r}}{{\partial R}}\frac{{\partial P_2}}{{\partial r}}\\
 + k_b ^2 \left( {\frac{{R_2 }}{{R_2  - R_1 }}} \right)^3 (R - R_1 )^2 P_2 = 0,
\end{array}
\end{equation}
By coordinate transformation in statement 6  in [15], the equation (108) is reduced to the 1-D sphere Bessel equation,
\begin{equation}  
\frac{\partial }{{\partial r}}r^2 \frac{{\partial P_2}}{{\partial r}} 
+ k_b ^2 r^2 P_2 = 0,  
\end{equation}  
Let $r = \frac{{R_2 }}{{R_2  - R_1 }}(R - R_1 )$, it is obvious that $P_2(R)$ in (104) can be translated to $P_2( r)$ 
\begin{equation}   
\begin{array}{l}
 P_2(R) =  - \frac{1}{{4\pi }}\frac{{e^{ik_b r_s } }}{{r_s }}j_0 \left( {k_b \frac{{R_2 }}{{R_2  - R_1 }}(R - R_1 )} \right) \\ 
  =  - \frac{1}{{4\pi }}\frac{{e^{ik_b r_s } }}{{r_s }}j_0 (k_b r)=P_2(r), \\ 
 \end{array}
\end{equation}  
so, the $P_2(r)$ in (110) is solution of the equation (99), therefore, $P_2( R) $ in (104)
\[
P_2(\vec R) =  - \frac{1}{{4\pi }}\frac{{e^{ik_b r_s } }}{{r_s }}j_0 \left( {k_b \frac{{R_2 }}{{R_2  - R_1 }}(R - R_1 )} \right),    (104)
\]
is solution of the acoustic equation (105), also satisfies interface boundary
pressure continuous condition (106)
\begin{equation}
\begin{array}{l}
 P_2( R^ - ) =P_2( R)|_{R =  R_1 }  \\ 
  =  - \frac{1}{{4\pi }}\frac{{e^{ik_b r_s } }}{{r_s }}j_0 \left( {k_b \frac{{R_2 }}{{R_2  - R_1 }}(R_1  - R_1 )} \right) 
  =  - \frac{1}{{4\pi }}\frac{{e^{ik_b r_s } }}{{r_s }} \\  
 \end{array}   
\end{equation} 
and velocity continuous interface condition (107) 
\begin{equation}
\begin{array}{l} 
\frac{{R_2 }}{{R_2  - R_1 }}(R - R_1 )^2 \frac{{\partial P_2}}{{\partial R}}|_{R = R_1 }  = \\
=\frac{{R_2 }}{{R_2  - R_1 }}(R - R_1 )^2 \frac{{\partial P_1}}{{\partial R}}|_{R = R_1 }  = 0, 
 \end{array}   
\end{equation}
Therefore there exist acoustic wave solution to satisfy the global acoustic 
equation system ((71) to (75) and (101) to (103)), the acoustic wave solution in (104) is propagation to penetrate into the inner sphere $R \le R_1$, the inner
sphere $R \le R_1$ is not cloaked. 
The induced anisotropic acoustic media (83),(84), (85) are same as (25),(24),(26),in the page 24301-3 in the paper [4], In statement 1, we proved that when  $j_1 (k_b R_1 ) \ne 0$, there exist no acoustic wave solution to
satisfy the global acoustic equation system with anisotropic the media in the paper [4], the anisotropic acoustic media (25),(24),(26),in page 24301-3 in [4]  is inconsistent with the background media,  the equation (27) in page 24301-3 in  the paper  [4]  is not satisfied and is contradiction equation   for $n=0$. The acoustic cloak in paper [4] is not acoustic no scattering cloak. In paper [4], authors write " Physically, this is because the radial mass density tends to infinity at the inner edge of the shell which reduces all radial particle motion to zero "  to explain their cloak, authors in [4] only explain the fluid velocity is zero at the inner edge of shell, but the pressure is nonzero and wave phase speed is nonzero in the edge of shell, moreover, we used same induced anisotropic acoustic media in paper [4], the radial mass density tends to infinity at the inner edge of the shell, we only add condition $j_1 (k_b R_1 ) = 0$, in satement 2, we proved that there exist nonzero acoustic wave propagation in (92) in the inner sphere,$R \le R_1$, the  inner sphere,$R \le R_1$, is not cloaked. Novely, we install the induced anisotropic acoustic media in paper [4] in the annular layer shell $R_1 \le R \le R_2$ and inner sphere $R \le R_1$, the radial mass density tends to infinity at the inner both edges of the shell, by the explain in paper [4] the  inner sphere should be cloaked, however, in statement 3, we proved that there exist nonzero acoustic wave in (104) is propagation in the inner sphere $R \le R_1$, the inner sphere $R \le R_1$, is
not cloaked. Therefore, The acoustic cloak in paper [4]
is not acoustic no scattering cloak. The $0$ to $R_1$  spherical radial linear transformation can not be used to induce acoustic no scattering cloak. Our proof are suitable for arbitrary $0$ to $R_1$  sphere radial transformation and its induced anisotropic, therefore the $0$ to $R_1$  spherical radial transformation can not be used to induce acoustic no scattering cloak [15]. 

\section {Conclusion}

Summary, if $j_1 (k_b R_1 ) \ne 0$, the transformation induced anisotropic acoustic media $(20) \ to \ (22)$ in the annular layer $R_1 \le R \le R_2$ are inconsistent with background medium in inner sphere $R\le R_1$, that caused there exsit no solution of the acoustic equations and the interface  continuous conditions equations system. if $j_1 (k_b R_1 ) = 0$, there exist acoustic wave solution to satisfy the above anisotropic acoustic equations and interface continuous condition equations system, the acoustic wave solution is propagation to penetrate into the inner sphere, the inner sphere can not be cloaked. If we install the transformation induced anisotropic acoustic media $(20) \ to \ (22)$ in the annular layer $R_1 \le R \le R_2$ and inner sphere $R \le R_1$, the acoustic wave solution is contnuous propagation to penetrate into the inner sphere, the inner sphere $R \le R_1$, can not be cloaked. Therefore, 0 to $R_1$ spherical radial transformation can not be used to induce acoustic no scattering cloak. The 0 to R1 spherical radial transformation can not be used to induce static electric conductivity no scattering cloak.
It is totally different from transformation, the GILD and GL no scattering modeling and inversion can be used to make GLHUA and GLLH double layer acoustic no scattering cloak.[1][2][5][7][8]         
\begin{acknowledgments}
We wish to acknowledge the support of the GL Geophysical Laboratory and thank the GLGEO Laboratory to approve the paper
publication. Authors thank to Professor P. D. Lax [9] for his concern and encouragements  Authors thank to Professor Michael Oristaglio and Professor You Zhong Guo for their encouragments
\end{acknowledgments}

\end{document}